\documentclass{aa}  

\usepackage{graphicx}
\usepackage{txfonts}
\usepackage{natbib}

\usepackage{bm}
\usepackage{esvect}
\usepackage{multirow}
\usepackage{varwidth}
\usepackage{array}

\renewcommand{\arraystretch}{1.2}
\DeclareMathOperator{\Tr}{Tr}
\begin{document}

   \title{Dynamical traceback age of the $\beta$~Pictoris moving group}

   \subtitle{}

   \author{N.~Miret-Roig\inst{1}
          \and
          P.A.B.~Galli\inst{1}
          \and
          W.~Brandner\inst{2}
          \and
          H. Bouy\inst{1}
          \and
          D. Barrado\inst{3}
          \and
          J. Olivares\inst{1}
          \and
          T. Antoja\inst{4}
          \and
          M. Romero-Gómez\inst{4}
          \and 
          F.~Figueras\inst{4}
          \and
          J. Lillo-Box\inst{3}
          }

   \institute{Laboratoire d'astrophysique de Bordeaux, Univ. Bordeaux, CNRS, B18N, allée Geoffroy Saint-Hilaire, 33615 Pessac, France.
         \email{nuria.miret-roig@u-bordeaux.fr}
        \and
            Max Planck Institute for Astronomy, Königstuhl 17, 69117, Heidelberg, Germany.
         \and
             Centro de Astrobiología (CSIC-INTA), ESAC Campus, Camino Bajo del Castillo s/n, 28692 Villanueva de la Cañada, Madrid, Spain.
        \and
            Institut de Ci\`{e}ncies del Cosmos, Universitat de Barcelona, IEEC, Mart\'{i} i Franqu\`{e}s 1, E08028 Barcelona, Spain
             }

   \date{Received ; accepted }

% \abstract{}{}{}{}{} 
% 5 {} token are mandatory
 
  \abstract
   {The $\beta$~Pictoris moving group is one of the most well-known young associations in the solar neighbourhood and several members are known to host circumstellar discs, planets, and comets. Measuring its age precisely is essential to the study of several astrophysical processes, such as planet formation and disc evolution, which are strongly age-dependent.}
   {We aim to determine a precise and accurate dynamical traceback age for the $\beta$~Pictoris moving group. }
    {Our sample combines the extremely precise \textit{Gaia} DR2 astrometry with ground-based radial velocities measured in an homogeneous manner. We use an updated version of our algorithm to determine dynamical ages. The new approach takes into account a robust estimate of the spatial and kinematic covariance matrices of the association to improve the sample selection process and to perform the traceback analysis. } 
   {We estimate a dynamical age of $18.5_{-2.4}^{+2.0}$~Myr for the $\beta$~Pictoris moving group. We investigated the spatial substructure of the association at the time of birth and we propose the existence of a core of stars that is more concentrated. 
   We also provide precise radial velocity measurements for 81 members of $\beta$~Pic, including ten stars with the first determinations of their radial velocities.}
   {Our dynamical traceback age is three times more precise than previous traceback age estimates and, more importantly, for the first time it reconciles the traceback age with the most recent estimates of other dynamical, lithium depletion boundaries and isochronal ages. This has been possible thanks to the excellent astrometric and spectroscopic precisions, the homogeneity of our sample, and the detailed analysis of binaries and membership. }
 

   \maketitle
%
%________________________________________________________________

% 
\section{Introduction}\label{sec:introduction}

Young local associations and moving groups are fundamental structures that help us to understand the stellar formation and evolution processes. They are small aggregates of stars (a few dozens) that share dynamical properties. For this reason, it is implicitly assumed that they were born at the same time and at the same place (from the same molecular cloud) and, therefore, they share the same chemical composition \citep{deZeeuw99, Jayawardhana00}. Most of the known associations are located nearby and allow for a detailed study of their properties.

One of the most well-known associations is $\beta$~Pictoris ($\beta$~Pic). It was discovered a couple of decades ago when \citet{Barrado99} identified the first two companions to the $\beta$~Pic star and \citet{Zuckerman+01} identified an additional set of 17 co-moving stars. Since then, many studies have contributed to increase the number of members of this association (e.g. \citealt{Torres+06}, \citealt{Malo2013}, \citealt{Binks15}, and \citealt{Riedel17}). Today, there are a few hundreds of candidate members of the $\beta$~Pic moving group, making it one of the richest associations. Its proximity ($\sim40$~pc) and observational characteristics (it is visible both from the southern and northern hemispheres) facilitated the discovery of members with a large diversity of stellar masses and very interesting properties, such as discs, confirmed exoplanets, and exocomets (\citealt{Kalas1995}, \citealt{Kalas2004}, \citealt{Lagrange2010}, \citealt{Lagrange2019}, \citealt{Chauvin2012}, and \citealt{Kiefer2014}).

The age is one of the most fundamental parameters to study stellar formation and evolution. $\beta$~Pic has an estimated age of $\sim$20~Myr (\citealt{Barrado99}, \citealt{Barrado2001}, \citealt{Mamajek14}, \citealt{Binks+14}), which is of particular interest for the study of several astrophysical processes, such as disc evolution and planet formation. However, different methods lead to a relatively broad range of values and errors ranging from $10.8$~Myr to $40$~Myr (see Table~\ref{tab:age_review} for a review of the literature age estimates of $\beta$~Pic).

Among the few techniques available to determine stellar ages, dynamical ages\footnote{The term "kinematical ages" is sometimes used for similar purposes.} have the advantage that they are independent of stellar evolutionary models. The main assumption of this method is that the stars were formed together, in the past, at a time when the association was most concentrated. This assumption is supported by the lithium and isochronal ages where there is no evidence of a significant age spread \citep{Mamajek14, Messina+16}. Several authors in the literature have used different techniques to traceback the positions and motions of the stars (linear trajectories, epicyclic approximation, orbital integration with a Galactic potential) and different definitions of the size of the association (e.g. standard deviation of the positions in a privileged direction, in 3D, maximum distance between members, pairwise encounters). Historically, the main limitations of the traceback analysis were the observational uncertainties in proper motions and the lack of trigonometric parallaxes and radial velocities to derive distances and spatial velocities \citep{Ortega02, Ortega04, Song+03}. After the \textit{Gaia} Data Release 2 \citep[DR2, ][]{GaiaCol+16,GaiaDR2} we have a large, uniform sample of stars with extremely precise parallax and proper motions. Additionally, several authors measured radial velocities of $\beta$~Pic members \citep[e.g. ][]{Torres+06, Shkolnik2012, Gagne2018b}, although these are highly inhomogeneous and systematic errors may exist between different studies. Currently, the main limitations of the traceback analysis are: 1) the availability of an homogeneous and precise dataset of radial velocities; 2) the design of a new strategy for the selection of kinematic members adequate for the  high-quality data at hand, and 3) a statistically robust approach for analysing the orbits and to establish a dynamical traceback age. In this study, we made a special effort to prepare a clean sample with precise and uniform data.

This work is structured as follows. In Section~\ref{sec:data}, we present the spectroscopic observations we carried out and the process for measuring precise radial velocities for new and archival data. We also describe the improvements of our method for selecting a bona fide sample of kinematic members from our initial list of candidates from the literature. In Section~\ref{sec:traceback}, we describe the algorithm used to derive the dynamical age and analyse in detail the orbits of the bona fide members. In Section~\ref{sec:discussion}, we discuss the results obtained and we present our conclusions in Sect.~\ref{sec:conclusions}.

\section{Data and sample selection}\label{sec:data}

\begin{table*}
   \begin{center}
    \caption{Number of sources at each step of the data selection process (see Sect.~\ref{sec:data}).}
    \label{tab:num_sources}
    \begin{tabular}{lr|r|c|c|c}
    \hline
    \hline
    \multicolumn{2}{l}{} &
    \multicolumn{1}{|c|}{\# Members} &
    \multicolumn{1}{c|}{ground-based RV error} &
    \multicolumn{1}{c|}{\textit{Gaia} RV error} & 
    \multicolumn{1}{c}{\textit{Gaia} $V_{tan}$ error}\\
   
    \multicolumn{2}{l}{} &
    \multicolumn{1}{|c|}{} &
    \multicolumn{1}{c|}{(km~s$^{-1}$)} &
    \multicolumn{1}{c|}{(km~s$^{-1}$)} &
    \multicolumn{1}{c}{(km~s$^{-1}$)}\\
    \hline
    \multicolumn{2}{l|}{Candidate members from literature}                         & 236 & 0.5$^*$ & 0.6~(55) & 0.19~(222) \\
    \hline                      
    \multicolumn{2}{l|}{High quality RV (this work)}                               &  81 & 0.3~~     & 0.6~(31) & 0.08~~~(79) \\
    \hline
    \multicolumn{2}{l|}{6D data (\textit{Gaia} astrometry + high quality RV)}      &  79 & 0.3~~     & 0.6~(31) & 0.08~~~~~~~~~ \\
    \multicolumn{1}{l}{\qquad Suspected SB (this work)} & \multicolumn{1}{r|}{2}   & & & &\\
    \multicolumn{1}{l}{\qquad SB from literature} &       \multicolumn{1}{r|}{35}  & & & &\\
    \multicolumn{1}{l}{\qquad Single} &                   \multicolumn{1}{r|}{42}  & & & &\\
    \hline 
    \multicolumn{2}{l|}{Single following kinematic criteria (see Sect.~\ref{subsec:vel_dist})}                      &  27 & \multirow{2}{*}{0.3~~} &  \multirow{2}{*}{0.6~(13)}  & \multirow{2}{*}{0.05~~~~~~~~~ } \\
    \multicolumn{2}{l|}{Confirmed by orbital analysis (bona fide sample)   }                         &  26 & & & \\
    \hline
    \hline
    \end{tabular}
    \end{center}{}
    \tablefoot{Columns 3--4 indicate the median radial velocity error obtained from ground-based surveys and from the \textit{Gaia} DR2 catalogue, respectively. For comparison, in column 5 we indicate the median tangential errors obtained with the \textit{Gaia} DR2 parallaxes and proper motions, obtained taking into account the correlation among the astrometric parameters. The number of \textit{Gaia} sources used to estimate the median velocity errors is indicated in brackets in each case.\\
    \noindent\tablefoottext{*}{median radial velocity errors published in the literature for the 137 stars with radial velocity previous to this work. This sample is inhomogeneous and may be affected by systematic errors among different studies.}} 

\end{table*}{}

In this section, we present a compilation of confirmed members and new candidates reported in the literature over the past decade. In order to have a sample with homogeneous stellar parameters, we use the 5D astrometric solution (positions, parallaxes, and proper motions) of the \textit{Gaia} DR2 catalogue. We complement these data with a set of radial velocities (from our own observations plus archival data) analysed using the same methodology. In this study we use the radial velocities published in the literature and in $Gaia$ DR2 only to compare with our own determinations. In Table~\ref{tab:num_sources}, we review the selection process from the initial compilation to the final sample.

Our initial sample is based on \citet{Torres2008}, \citet{Schlieder2012},  \citet{Malo2013}, \citet{Malo2014}, \citet{Gagne2015b,Gagne2015a}, \citet{Alonso-Floriano2015}, \citet{Messina2017}, \citet{Gagne2018a} and \citet{Gagne2018b}. This results in a sample of 236~stars after removing the sources in common between the studies. Binaries and multiple systems are counted as one single object unless they have been resolved in previous studies. These authors used different algorithms based on the kinematics (and included the photometry in some cases) to identify new candidates or confirm members of $\beta$~Pic. Most of these studies are pre-\textit{Gaia} or were carried with partial information (missing parallaxes or radial velocities). For this reason, it has been necessary to develop a tool to reject kinematic outliers with our homogeneous and precise astrometry and spectroscopy (see Sect.~\ref{subsec:vel_dist}).

\subsection{Proper motions and parallaxes}

We use the proper motions and parallaxes of the \textit{Gaia} DR2 catalogue which constitute the most recent and precise astrometric measurements available to date for our sample. To identify the \textit{Gaia} DR2 counterparts of the stars in our sample we used the 2MASS source identifier (which are given in the original tables used to construct our initial sample) and the \texttt{TMASS\_BEST\_NEIGHBOUR} table available in the \textit{Gaia} archive. For 42 sources we did not find a counter part with this procedure, so we manually refine the match considering position and magnitude. Finally, we find proper motions and parallaxes for 222~stars in our initial sample. There are eight sources in $Gaia$ DR2 with only the two-parameter solution and six that are not in $Gaia$ DR2 (see App.~\ref{App:match-Gaia}). The median of the uncertainties of this sample are $\sim0.1$~mas~yr$^{-1}$ in proper motions and 0.08~mas in parallax which lead to a median error in tangential velocity of 0.19~km~s$^{-1}$, obtained by taking into account the correlations among the astrometric parameters (see Table~\ref{tab:num_sources}).

\subsection{Radial velocities}

The scarcity and quality of the radial velocities of $\beta$~Pic stars are currently two of the main limitations for deriving an accurate estimate of the dynamical age of the association.
Even though many radial velocity measurements are available in the literature (e.g. \citealt{Torres+06}, \citealt{Kharchenko2007}, \citealt{Shkolnik2012}, \citealt{Elliott+2014}, \citealt{Gagne2018b}), we resort to a re-analysis of the spectra available in public archives to ensure that all the radial velocities are derived using the same methodology. The consistency and homogeneity of the individual measurements is indeed particularly important in a dynamical traceback analysis \citep[see e.g.][]{Miret-Roig+18}. 

\subsubsection{New spectroscopic observations}\label{subsec:new-spec}

\begin{table} 
\centering
\caption{Spectra analysed in this study from our spectroscopic observations plus archival data. 
\label{tab:num-spectra}}
\begin{tabular}{lrr@{$-$}rr@{/}l}
\hline\hline
  \multicolumn{1}{c}{Spectrograph} &
  \multicolumn{1}{c}{$R$} & 
  \multicolumn{2}{c}{$\Delta\lambda$} & 
  \multicolumn{2}{c}{\# Spectra} \\
  
  \multicolumn{1}{c}{} &
  \multicolumn{1}{c}{} & 
  \multicolumn{2}{c}{(nm)} & 
  \multicolumn{2}{c}{total/this work} \\
\hline\hline
FEROS  &  $48\,000$ & $350$\, & $920$  & 167\,\, & \,\,45  \\
HARPS  & $115\,000$ & $378$\, & $691$  & 138\,\, & \,\,\, 0 \\
ELODIE &  $45\,000$ & $385$\, & $680$  &  45\,\, & \,\,\, 0 \\
SOPHIE &  $75\,000$ & $387$\, & $694$  &  62\,\, & \,\,62  \\
UVES   & $110\,000$ & $300$\, & $1100$ & 277\,\, & \,\,\, 0 \\
CAFE   &  $62\,000$ & $407$\, & $925$  &  34\,\, & \,\,34  \\
\hline\hline
\end{tabular}
\tablefoot{Number of spectra analysed and number of new spectra obtained in this study with different instruments. The total number of spectra analysed is 723 and 141 of them are new. The (maximum) resolving power and spectral range of each spectrograph are indicated.}
\end{table}

We performed spectroscopic observations of $\beta$~Pic stars with three different instruments. The FEROS spectrograph \citep{Kaufer1999} mounted on the ESO/MPG 2.2~m telescope operated at La Silla (Chile) was used to collect the spectra of 43~stars as part of programme 103.A-9009 (PI: W.~Brandner). These observations were performed in OBJCAL mode that allows for simultaneous acquisition of the target spectrum and the calibration lamp during July and August 2019. We observed 8~stars with the CAFE spectrograph  \citep{Aceituno+2013,Lillo-Box+2020} mounted on the 2.2~m telescope of the Calar Alto Observatory (programme: H18-2.2-015, F19-2.2-002, PI: D.~Barrado). The observations were carried out from July to October 2018, right after the upgrade of the instrument. The data were processed using the new instrument pipeline described in \citet{Lillo-Box+2020}, which performs the basic reduction and extracts the radial velocities. Finally, another 14~stars were observed with the SOPHIE spectrograph \citep{Perruchot+2008} mounted on the 1.93~m telescope of the Haute-Provence Observatory (programmes: 2018A-PNPS005, 2019A-PNPS008, PI: H.~Bouy). These spectra were obtained in August 2018 and May 2019 and were processed with the instrument standard data reduction pipeline which measures radial velocities by numerical cross-correlation techniques. The median signal-to-noise ratio (S/N) of our observations is 25.

\subsubsection{Spectroscopic archival data}

In addition to the observations conducted by our team we did an exhaustive search for the spectra available in public archives. As shown in Table~\ref{tab:num-spectra}, a total of 582 spectra have been collected from the European Southern Observatory (ESO) and the ELODIE archives. We reanalysed all these data (see Sect.~ \ref{section2.3.2}) and provide radial velocities for a larger number of stars.
Table~\ref{tab:num-spectra} shows the instruments that have been used in this study and the respective number of spectra analysed in each case. We specify the number of new spectroscopic observations presented in this work (see Sect.~\ref{subsec:new-spec}) which constitute a $20\%$ of all the spectra.  We note that some sources have been observed several times with the same or various instruments. In fact, the 723 spectra correspond to 81 different stars, 54\% of which have been observed once, 18\% twice, and the rest three or more times. We refer to Sect.~\ref{section2.3.2} for a description of how we combined the different radial velocity measures for the same star.

\subsubsection{Radial velocities determination}\label{section2.3.2}

The observed and downloaded spectra were reduced using the official pipeline available for each instrument. We derived radial velocities by cross-correlating the reduced spectra of the stars with the closest mask to its spectral type. We used six different masks of spectral types A0, F0, G2, K0, K5, and M5, along with the iSpec routines for this purpose \citep{Blanco-Cuaresma2014,Blanco-Cuaresma2019}. This procedure follows the cross-correlation technique \citep{Baranne1996,Pepe2002} and fits a Gaussian profile to the cross-correlation function (CCF) to derive the radial velocity and associated uncertainty. We discard the radial velocity measurements resulting from a poor fit to the CCF due to, for example, a low S/N of the spectrum or a mismatch between the spectral type of the star and the adopted mask. We used the effective temperatures given in \textit{Gaia} DR2 as a rough estimate of the spectral type of the star to choose the corresponding mask. For each star, we compute the radial velocity scatter from the results obtained with three different masks: the closest mask (M) to the spectral type of the star, one before (M-1), and one after (M+1). We add this number in quadrature to the formal uncertainty returned from the iSpec routines. The later step accounts for the observed fluctuation on the radial velocity results derived from different masks\footnote{This method provided an overestimate uncertainty for the $\beta$~Pic star since it is a fast rotator ($v\sin i =120$~km~s~$^{-1}$, \citealt{Lagrange2019}) and only the A0 mask provides reasonable CCF fit. The formal error returned by the iSpec routines is 2.2~km~s~$^{-1}$, a 60\% smaller than the final uncertainty we obtain from different masks (5.5~km~s~$^{-1}$).}.

We derive radial velocities for 81 stars of our initial sample of $\beta$~Pic candidates by combining our own observations with archival spectra (see Table available at CDS). In the case of multiple radial velocity measurements for the same star, we proceed as follows. For each radial velocity solution (for a given star), we generate a sample of 10\,000 synthetic measurements from a Gaussian distribution where the mean and variance correspond to the radial velocity and its uncertainty. We repeat this process for all radial velocity measurements of the star. Then we take the mean of the joint distribution of synthetic radial velocities as our final result for the radial velocity of the star. The uncertainties on the resulting radial velocity are computed from the 16\% and 84\% percentiles of the joint distribution of synthetic radial velocities. We note that for ten of the stars (12\% of the radial velocities we determine), our radial velocity is the first measurement ever taken. This is an important product of our work since these data can be used to assess the membership and to study the dynamics of the association in 6D. Additionally, six of them are in our final bona fide sample of 26 stars (see Sect~\ref{subsec:vel_dist}). 

In Table~\ref{tab:num_sources}, we compare the quality of our radial velocities with the \textit{Gaia} DR2 catalogue and with previous ground-based spectroscopic surveys. We find a radial velocity in the literature for 137 sources in our initial sample. These measurements come from a variety of different surveys with different qualities and methods to determine the radial velocity. Our measurements are homogeneous and about 40\% more precise than this compilation which is crucial for the success of our work. We see that the radial velocities of our sample are twice as precise as the \textit{Gaia} DR2 radial velocities and we have a measurement for a larger number of sources. We identified and discarded 35 sources which have been classified as binaries in previous works. In order to include the binaries in our study, we would need to determine the radial velocity motion of the centre of mass and that is beyond the scope of this work.

\begin{figure}
\begin{center}
\includegraphics[width = \columnwidth]{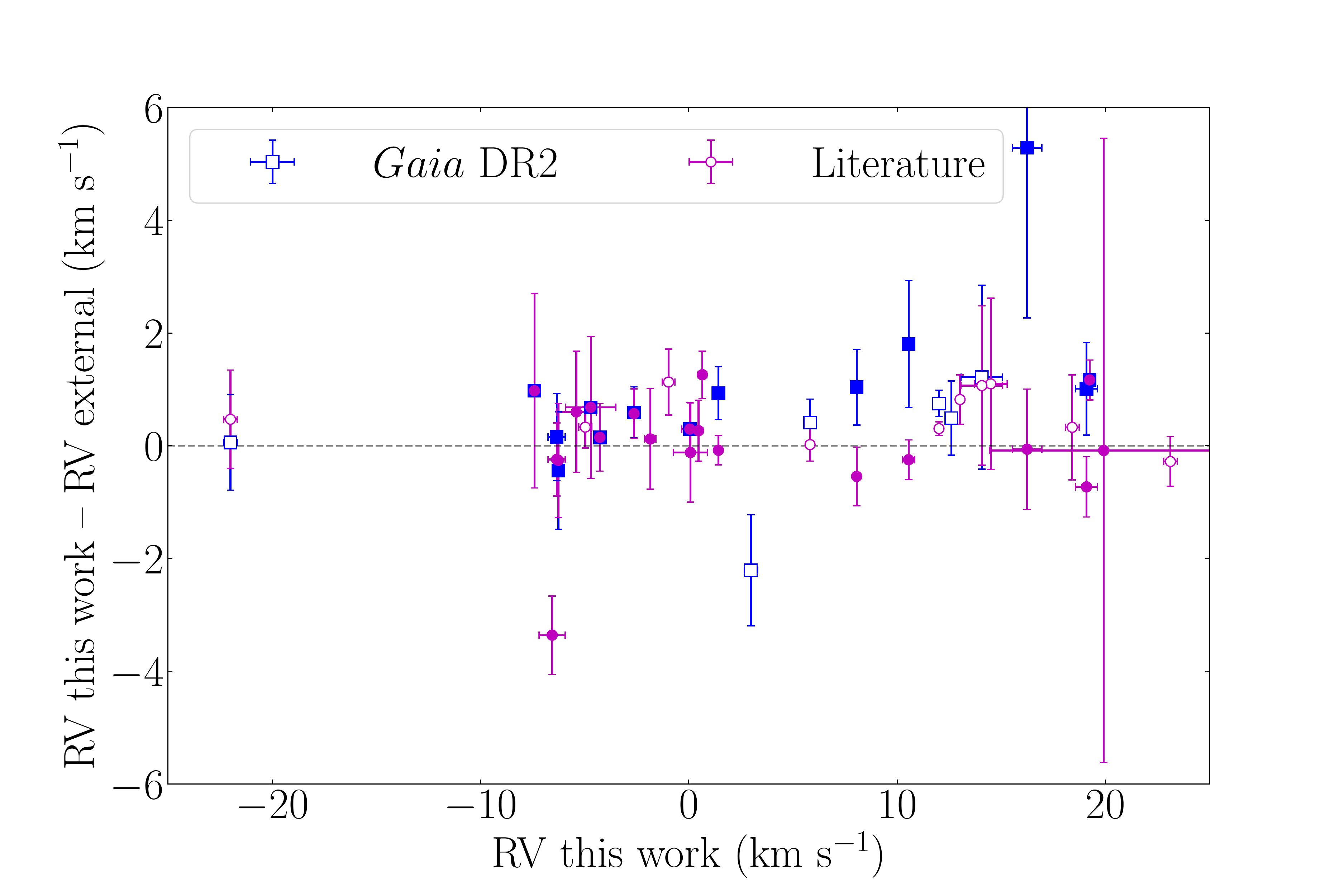}
\caption{
\label{RV_comp}
Radial velocity residuals between this work and external measures: \textit{Gaia} DR2 (blue squares) and previous spectroscopic measures in the literature (magenta open circles). For this comparison, we used the 42 single stars with 6D data (see Table~\ref{tab:num_sources}). Our final sample of 26 stars is represented by the filled markers. We note that our radial velocity uncertainties (horizontal error bars) are smaller than the markers in most of the cases. The largest uncertainty corresponds to the $\beta$~Pic star (see footnote~2).}
\end{center}
\end{figure}

Figure~\ref{RV_comp} shows the comparison of the radial velocities derived in our study with the ones in \textit{Gaia} DR2 and the ones in other spectroscopic surveys in the literature for the 42 single stars with 6D data in our sample. We found hints of binarity in two sources (\object{2MASS J19312434-2134226} and \object{2MASS J22571130+3639451}) and we discarded them from the analysis (see App.~\ref{App:discarded}). The median difference and root mean square error (RMSE) between the \textit{Gaia} DR2 radial velocities and our measurements are 0.7~km~s$^{-1}$ and 1.0~km~s$^{-1}$, respectively. These values are computed while disregarding the source with a radial velocity difference of about 5~km~s$^{-1}$. The \textit{Gaia} DR2 radial velocity of this star is based only in two transits, which is probably the reason for its large uncertainty. If we compare the radial velocities from the literature and our sample, we obtain a median difference and RMSE of 0.3~km~s$^{-1}$ and 0.9~km~s$^{-1}$, respectively. Since we believe that the homogeneity and precision of our radial velocities is superior to any other sample, we only use our measurements in the current analysis. 

\subsection{Kinematic sample selection}\label{subsec:vel_dist}

\begin{sidewaystable*}
    \centering
    \caption{Final bona fide sample of 26 members of $\beta$~Pic selected to determine the dynamical age. }
    \label{tab:26_members}

\begin{tabular}{|l|r|r|r|r|r|r|r|r|r|c|}
\hline
\hline
  \multicolumn{1}{|c|}{2MASS ID} &
  \multicolumn{1}{c|}{SpT\tablefootmark{a} (ref)} &
  \multicolumn{1}{c|}{T$_\textup{eff, GDR2}$ (K)} &
  \multicolumn{1}{c|}{$G$ (mag)} &
  \multicolumn{1}{c|}{$X$ (pc)} &
  \multicolumn{1}{c|}{$Y$ (pc)} &
  \multicolumn{1}{c|}{$Z$ (pc)} &
  \multicolumn{1}{c|}{$U$ (km~s$^{-1}$)} &
  \multicolumn{1}{c|}{$V$ (km~s$^{-1}$)} &
  \multicolumn{1}{c|}{$W$ (km~s$^{-1}$)} &
  \multicolumn{1}{c|}{Core}\\
\hline
J00172353-6645124 & M3.0V (1) & 3630 & 11.3 & 14.60 $\pm$ 0.02 & -18.59 $\pm$ 0.02 & -28.21 $\pm$ 0.04 & -10.51 $\pm$ 0.12 & -16.09 $\pm$ 0.15 &  -8.60 $\pm$ 0.22 & Y\\
J04593483+0147007 & M0V   (2) & 3986 &  9.3 &-21.28 $\pm$ 0.02 &  -6.77 $\pm$ 0.01 &  -9.83 $\pm$ 0.01 & -12.42 $\pm$ 0.46 & -16.30 $\pm$ 0.15 &  -9.28 $\pm$ 0.21 & Y\\
J05004714-5715255 & M0V   (2) & 4033 &  9.4 & -1.54 $\pm$ 0.00 & -21.32 $\pm$ 0.02 & -16.33 $\pm$ 0.01 & -11.17 $\pm$ 0.02 & -16.50 $\pm$ 0.15 &  -9.09 $\pm$ 0.12 & Y\\
J05471708-5103594 & A6V   (3) & 7100 &  3.7 & -3.43 $\pm$ 0.02 & -16.65 $\pm$ 0.11 & -10.06 $\pm$ 0.07 & -11.07 $\pm$ 0.96 & -15.79 $\pm$ 4.63 &  -9.21 $\pm$ 2.80 & Y\\
J06182824-7202416 & K4V   (2) & 4555 &  9.3 &  7.59 $\pm$ 0.01 & -33.75 $\pm$ 0.03 & -18.58 $\pm$ 0.02 & -10.50 $\pm$ 0.14 & -16.46 $\pm$ 0.61 &  -8.71 $\pm$ 0.33 & Y\\
J16572029-5343316 & M3    (4) & 3612 & 11.3 & 45.37 $\pm$ 0.28 & -21.65 $\pm$ 0.13 &  -5.88 $\pm$ 0.04 &  -7.35 $\pm$ 0.16 & -15.87 $\pm$ 0.13 & -10.49 $\pm$ 0.08 & N\\
J17020937-6734447 & (M4)  (5) & 3712 & 12.8 & 32.00 $\pm$ 0.05 & -23.67 $\pm$ 0.04 & -10.97 $\pm$ 0.02 &  -8.20 $\pm$ 0.72 & -16.62 $\pm$ 0.53 &  -9.09 $\pm$ 0.25 & N\\
J17024014-4521587 & (M2)  (5) & 3914 & 10.7 & 30.33 $\pm$ 0.06 & -10.06 $\pm$ 0.02 &  -1.24 $\pm$ 0.00 &  -8.67 $\pm$ 0.17 & -16.51 $\pm$ 0.07 & -10.21 $\pm$ 0.03 & Y\\
J17444256-5315471 & (M3)  (5) & 3866 & 12.9 & 48.62 $\pm$ 0.17 & -18.83 $\pm$ 0.07 & -11.34 $\pm$ 0.04 &  -7.17 $\pm$ 0.35 & -16.93 $\pm$ 0.15 & -10.12 $\pm$ 0.09 & N\\
J17483374-5306118 & (M2)  (5) & 3962 & 12.9 & 70.36 $\pm$ 0.29 & -26.64 $\pm$ 0.11 & -16.96 $\pm$ 0.07 &  -7.44 $\pm$ 0.12 & -16.81 $\pm$ 0.09 &  -9.20 $\pm$ 0.06 & Y\\
J18041617-3018280 & (M2)  (5) & 3814 & 11.7 & 54.93 $\pm$ 0.15 &   0.82 $\pm$ 0.00 & - 4.03 $\pm$ 0.01 &  -7.83 $\pm$ 0.24 & -14.52 $\pm$ 0.05 &  -8.45 $\pm$ 0.04 & Y\\
J18055491-5704307 & (M2)  (5) & 3884 & 12.4 & 49.55 $\pm$ 0.16 & -21.31 $\pm$ 0.07 & -16.18 $\pm$ 0.05 &  -8.58 $\pm$ 0.19 & -15.54 $\pm$ 0.10 &  -8.09 $\pm$ 0.07 & Y\\
J18092970-5430532 & (M4)  (5) & 3826 & 13.4 & 34.91 $\pm$ 0.11 & -13.14 $\pm$ 0.04 & -10.76 $\pm$ 0.03 & -10.23 $\pm$ 0.24 & -15.13 $\pm$ 0.11 &  -8.05 $\pm$ 0.08 & N\\
J18161236-5844055 & (M3)  (5) & 3563 & 11.5 & 26.21 $\pm$ 0.05 & -11.84 $\pm$ 0.02 &  -9.68 $\pm$ 0.02 &  -7.78 $\pm$ 0.35 & -17.09 $\pm$ 0.16 & -10.26 $\pm$ 0.13 & N\\
J18281651-4421477 & (M2)  (5) & 3996 & 12.6 & 77.46 $\pm$ 0.28 & -13.26 $\pm$ 0.05 & -20.69 $\pm$ 0.07 &  -6.84 $\pm$ 0.22 & -16.37 $\pm$ 0.08 &  -8.99 $\pm$ 0.08 & Y\\
J18283524-4457280 & (K7)  (5) & 4190 & 11.6 & 79.32 $\pm$ 0.20 & -14.36 $\pm$ 0.04 & -21.66 $\pm$ 0.06 &  -6.45 $\pm$ 0.33 & -16.22 $\pm$ 0.08 &  -8.95 $\pm$ 0.10 & Y\\
J18420694-5554254 & M3.0V (1) & 3753 & 12.4 & 45.01 $\pm$ 0.12 & -16.58 $\pm$ 0.04 & -18.35 $\pm$ 0.05 &  -8.52 $\pm$ 0.19 & -15.40 $\pm$ 0.08 &  -8.28 $\pm$ 0.08 & Y\\
J19225894-5432170 & F6V   (2) & 6437 &  6.9 & 41.30 $\pm$ 0.11 & -12.80 $\pm$ 0.04 & -21.32 $\pm$ 0.06 &  -8.89 $\pm$ 0.34 & -15.43 $\pm$ 0.11 &  -8.06 $\pm$ 0.18 & Y\\
J19233820-4606316 & M0    (4) & 4050 & 11.2 & 64.02 $\pm$ 0.19 &  -9.10 $\pm$ 0.03 & -29.50 $\pm$ 0.09 &  -6.73 $\pm$ 0.74 & -16.37 $\pm$ 0.12 &  -9.76 $\pm$ 0.35 & N\\
J19243494-3442392 & M4.0V (1) & 4045 & 12.8 & 48.13 $\pm$ 0.22 &   3.22 $\pm$ 0.01 & -18.92 $\pm$ 0.09 &  -9.47 $\pm$ 0.58 & -16.14 $\pm$ 0.09 &  -8.88 $\pm$ 0.24 & N\\
J19481651-2720319 & (M2)  (5) & 3944 & 12.2 & 57.78 $\pm$ 0.22 &  13.52 $\pm$ 0.05 & -26.17 $\pm$ 0.10 &  -7.69 $\pm$ 0.15 & -14.99 $\pm$ 0.06 &  -9.20 $\pm$ 0.08 & Y\\
J20013718-3313139 & M1    (4) & 3938 & 11.5 & 52.34 $\pm$ 0.13 &   7.26 $\pm$ 0.02 & -28.47 $\pm$ 0.07 &  -7.68 $\pm$ 0.16 & -15.78 $\pm$ 0.04 &  -9.13 $\pm$ 0.10 & N\\
J20090521-2613265 & F5V   (6) & 6450 &  7.1 & 42.62 $\pm$ 0.09 &  12.23 $\pm$ 0.03 & -23.41 $\pm$ 0.05 &  -6.96 $\pm$ 1.02 & -14.69 $\pm$ 0.29 & -10.27 $\pm$ 0.56 & N\\
J20333759-2556521 & M4.5V (4) & 3864 & 13.1 & 34.55 $\pm$ 0.17 &  11.39 $\pm$ 0.06 & -23.62 $\pm$ 0.12 &  -8.44 $\pm$ 0.76 & -14.85 $\pm$ 0.26 &  -9.59 $\pm$ 0.52 & Y\\
J21100535-1919573 & M2    (4) & 3770 & 10.8 & 21.87 $\pm$ 0.06 &  12.25 $\pm$ 0.04 & -20.26 $\pm$ 0.06 &  -9.91 $\pm$ 0.28 & -15.17 $\pm$ 0.16 &  -9.78 $\pm$ 0.26 & Y\\
J22424896-7142211 & K7V   (2) & 4065 &  9.8 & 19.57 $\pm$ 0.02 & -18.94 $\pm$ 0.02 & -24.54 $\pm$ 0.02 & -10.25 $\pm$ 0.08 & -15.84 $\pm$ 0.08 &  -7.98 $\pm$ 0.10 & Y\\
\hline
\hline
\end{tabular}
\tablebib{ (1)~\citet{Riedel+2017b}; (2)~\citet{Torres+06}; (3)~\citet{Gray+2006}; (4)~\citet{Riaz+2006}; (5)~\citet{Gagne2018b}; (6)~\citet{Houk1982}}
\tablefoot{
\tablefoottext{a}{Spectral types between parentheses were estimated from the absolute $Gaia~G-$band magnitude.}
}
\end{sidewaystable*}{}

\begin{figure}
    \includegraphics[width = \columnwidth]{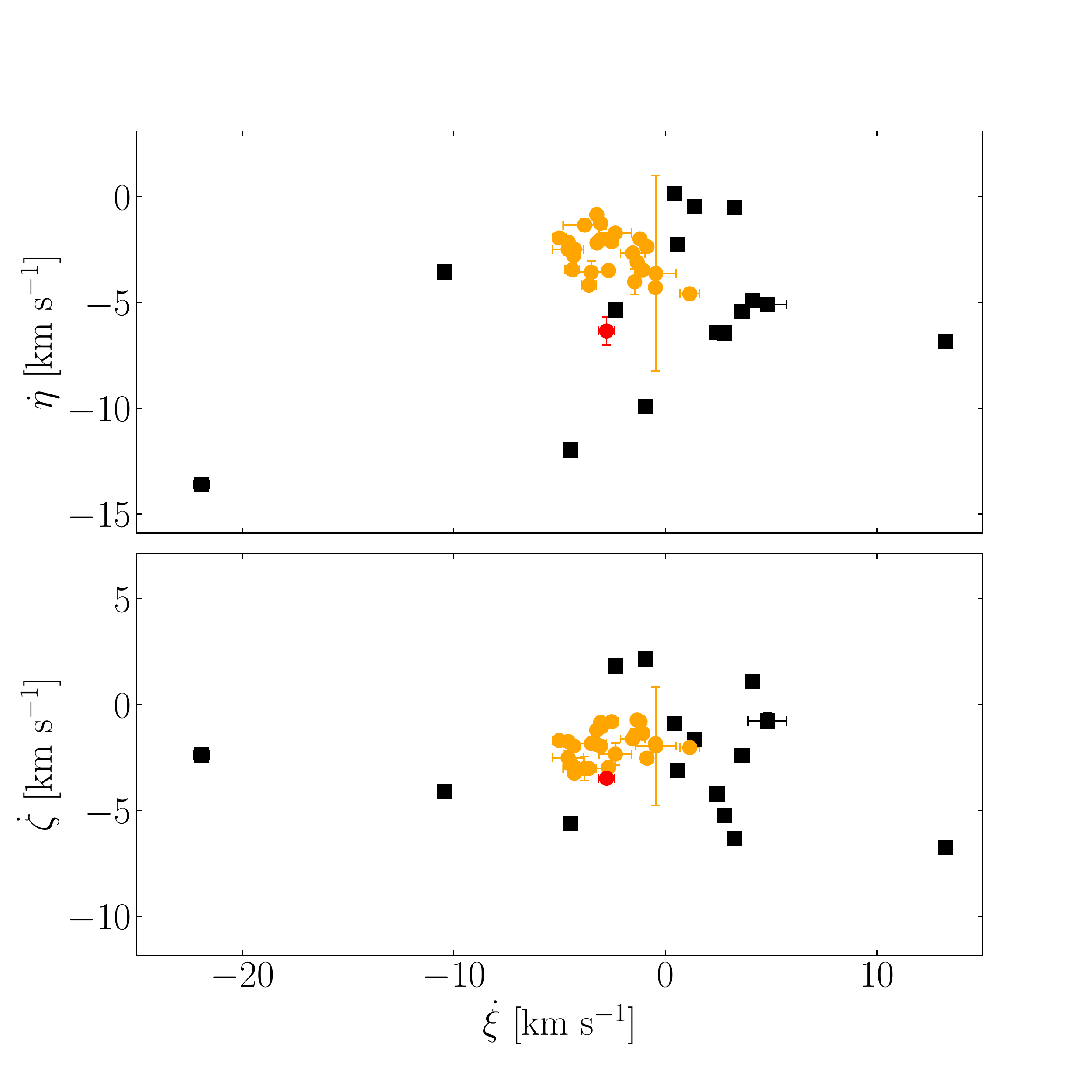}
    \caption{Present velocity distribution in the Galactic plane (top) and in the vertical plane (bottom) of the sample of 42 single sources with 6D data. The sample of 26 kinematically selected members is represented by the orange dots and the kinematically rejected sources are the black squares. The source \object{2MASS J11493184--7851011} (red dot) is retained by our kinematical criteria but is discarded due to its orbit (see text).}
    \label{fig:vel_present}
\end{figure}{}

In this section, we present the kinematic selection that we designed to discard kinematic outliers in our sample. Kinematic outliers in the context of the present study refer to sources with a velocity significantly different than the group, either because they are non-members or because they are variable due to multiplicity, for example. First, we introduce the notation we adopted to refer to position and velocity coordinate systems. We used the curvilinear heliocentric coordinates ($\xi^\prime$, $\eta^\prime$, $\zeta^\prime$) defined in \citet{Asiain99} to place the stars in the configuration space. This coordinate system is centred at the current position of the Sun ($R_\odot = 8.4$~kpc) and rotates around the Galactic centre with a frequency of the circular velocity of $\omega_\odot = 28.81$~km~s$^{-1}$~kpc$^{-1}$ \citep{Irrgang+13}. It has the advantage that it minimises the variation in each component of the configuration space. The radial component $\xi^\prime$ points towards the Galactic anti-centre, the azimuthal component $\eta^\prime$ is measured along the circle of radius $R_\odot$ and is positive in the sense of the galactic rotation, and the vertical component  $\zeta^\prime$ is defined positive towards the north Galactic pole. We also refer to the corresponding velocities as $\dot{\xi}^\prime, \dot{\eta}^\prime, \dot{\zeta}^\prime$. The second reference system considered in this work is the Cartesian heliocentric system. The spatial components $X, Y, Z,$ along with the velocity components $U, V, W$, are defined with $X, U$ pointing towards the Galactic centre, $Y, V$ towards the direction of Galactic rotation, and $Z, W$ towards the north Galactic pole. We use a peculiar solar motion of $(U_\odot, V_\odot, W_\odot) = (11.1,12.24,7.25)$~km~s$^{-1}$ \citep{Schonrich10}. 

In Figure~\ref{fig:vel_present} we represent the velocity distribution of the 42 single stars with \textit{Gaia} astrometry and radial velocities from this work. We see that a number of sources have a significant scatter. Most of them were classified as members of $\beta$~Pic with pre-\textit{Gaia} astrometry or with no radial velocity information and clearly appear to be kinematic outliers with our extremely precise data.
We discard the kinematic outliers in the 3D velocity distribution $(\dot{\xi}^\prime, \dot{\eta}^\prime, \dot{\zeta}^\prime)$ in a similar way to what we did in \citet{Miret-Roig+18}. The major improvement is that in this work we use a robust estimator of the covariance matrix \citep[the Minimum Covariance Determinant from Sklearn,][]{scikit-learn} to fit the central location ($\vv{\bm{\mu}}$) and the covariance matrix ($\mathbf{\Sigma}$) of the velocity ellipsoid of the association. Then, we compute the Mahalanobis distance of each object defined as:
\begin{equation}
    D_M(\vv{\bm{x}}) = \sqrt{(\vv{\bm{x}}-\vv{\bm{\mu}})^T \, \mathbf{\Sigma}^{-1} \, (\vv{\bm{x}}-\vv{\bm{\mu}})}.
    \label{eq:mahalanobis_dist}
\end{equation}{}
In Figure~\ref{fig:mahalanobis_dist_kin_selection}, we show the distribution of Mahalanobis distances. We use the percentile $p_{65}$ to discard the kinematic outliers and retain 27 kinematic members (dots in Fig.~\ref{fig:vel_present}). This threshold is empiric and represents the best compromise between rejecting kinematic outliers which hinder the traceback analysis and keeping kinematic members in the final sample. When we compute the orbits of our targets (see Sect.~\ref{subsec:DA}) we immediately see that one of them (2MASS J11493184--7851011, red circle in Fig.~\ref{fig:vel_present}) has an orbit that is significantly different from the main group and thus, we discard this object. This star has a kinematics similar to $\beta$~Pic but it is at $>3\sigma$ in positions with respect to $\beta$~Pic. We also checked that this object has the largest Mahalanobis distance to the centre of the velocity distribution. We refer to App.~\ref{App:discarded} for a detailed discussion, source-by-source, of the kinematically rejected sources. 
The final sample contains 26 bona fide members of $\beta$~Pic and their 3D positions and velocities are given in Table~\ref{tab:26_members} (available at CDS).

\subsection{Bona fide $\beta$ Pic sample}

\setlength{\tabcolsep}{7pt} 
\renewcommand{\arraystretch}{1.5} 
\begin{table*}[]
    \centering
    \caption{Parameters of the distribution in positions (in pc) and in velocities (in km~s$^{-1}$) of the 26 bona fide kinematic members of $\beta$~Pic in the present ($t=0$~Myr). }
    \begin{tabular}{ccc|ccc|ccc|ccc}
    \hline
    \hline
   \multicolumn{12}{c}{Positions}\\
    \hline
    \hline
    $X$ & $Y$ & $Z$ & $\sigma_{obs, X}$ & $\sigma_{obs, Y}$ & $\sigma_{obs, Z}$ & $\sigma_{err, X}$ & $\sigma_{err, Y}$ & $\sigma_{err, Z}$ & $\sigma_{int, X}$ & $\sigma_{int, Y}$ & $\sigma_{int, Z}$\\
    \hline
   $47.49$ & $-7.89$ & $-17.92$ & $16.04$ & $13.18$ & $7.44$ & $0.11$ & $0.04$ & $0.05$ & $15.93$ & $13.14$ & $7.39$\\
   \hline
   \hline
   \multicolumn{12}{c}{Velocities}\\
    \hline
    \hline
    $U$ & $V$ & $W$ & $\sigma_{obs, U}$ & $\sigma_{obs, V}$ & $\sigma_{obs, W}$ & $\sigma_{err, U}$ & $\sigma_{err, V}$ & $\sigma_{err, W}$ & $\sigma_{int, U}$ & $\sigma_{int, V}$ & $\sigma_{int, W}$\\
    \hline
   $-8.74$ & $-16.16$ & $-9.98$ & $1.49$ & $0.54$ & $0.70$ & $0.24$ & $0.11$ & $0.11$ & $1.25$ & $0.43$ & $0.59$\\
    \hline
    \hline
    \end{tabular}
    \tablefoot{Columns indicate: (1--3) central location of the distribution, (4--6) robust standard deviation, (7--9) median errors, and (10--12) a rough estimate of the intrinsic dispersion, computed as $\sigma_{int}^2 = \sigma_{obs}^2 - \sigma_{err}^2$.}

    \label{tab:median_velocity}
\end{table*}

In this paper, we make a substantial effort to prepare a robust sample of $\beta$~Pic members with the best precision possible in their determination of the positions in the 6D space phase. Then we used this valuable data to identify and remove kinematic outliers. In this section, we review the main characteristics of our final sample.

The relative error in the parallax of these members is less than 1\% which allows us to compute the distance as the inverse of parallax. We note that four stars have a parallax error $<0.1\%$ at distances up to 50~pc. The median relative errors in proper motions are of 0.3\% in right ascension ($\mu_{\alpha^*}$) and 0.09\% in declination ($\mu_\delta$). The precision in $\mu_{\alpha^*}$ and $\mu_\delta$ is similar but a few members have $\mu_{\alpha^*}$ close to zero which increases the relative error.
The $\beta$~Pic star is the brightest source ($G=3.7$~mag) and causes a fraction of the pixel used in the standard \textit{Gaia} DR2 analysis to be saturated. Hence, measurements of its centroid position and the resulting astrometry are less precise than for fainter sources ($G> 6$~mag) \citep{Lindegren+2018}.

In Table~\ref{tab:median_velocity}, we provide the median heliocentric position and velocity of $\beta$~Pic. We see that the observational uncertainties in positions ($\sigma_{err}$) are of the order of tenths of parsecs and thus, the observed dispersion ($\sigma_{obs}$) can be interpreted as an intrinsic dispersion ($\sigma_{int}$). The dispersion in the Galactic plane ($X, Y$ components) is twice the vertical dispersion ($Z$). When we look at the velocity dispersion, we find that the median errors in velocity ($\sigma_{err}$) are significantly smaller than the velocity dispersion observed ($\sigma_{obs}$), indicating the presence of an intrinsic cosmic dispersion ($\sigma_{int}$). Therefore, the dispersion we observe in Fig.~\ref{fig:vel_present} is intrinsic and not due to observational errors. The velocity ellipsoid is elongated in the radial direction (towards the Galactic centre) with a dispersion that is twice that of those in the other two directions. The typical velocity dispersions observed in molecular clouds are of the order of 0.5 to 1~km~s$^{-1}$ in nearby, low-mass star-forming regions \citep[and references therein]{Hennebelle+2012, Heyer+2015}, similar to the velocity dispersion we find in $\beta$~Pic.

\section{Traceback analysis}\label{sec:traceback}

In this section, we describe our methodology to perform the traceback analysis which is based on the work of \citet{Miret-Roig+18}, with some improvements.

\subsection{Towards a dynamical age estimate}\label{subsec:DA}

\begin{figure}
    \includegraphics[width = \columnwidth]{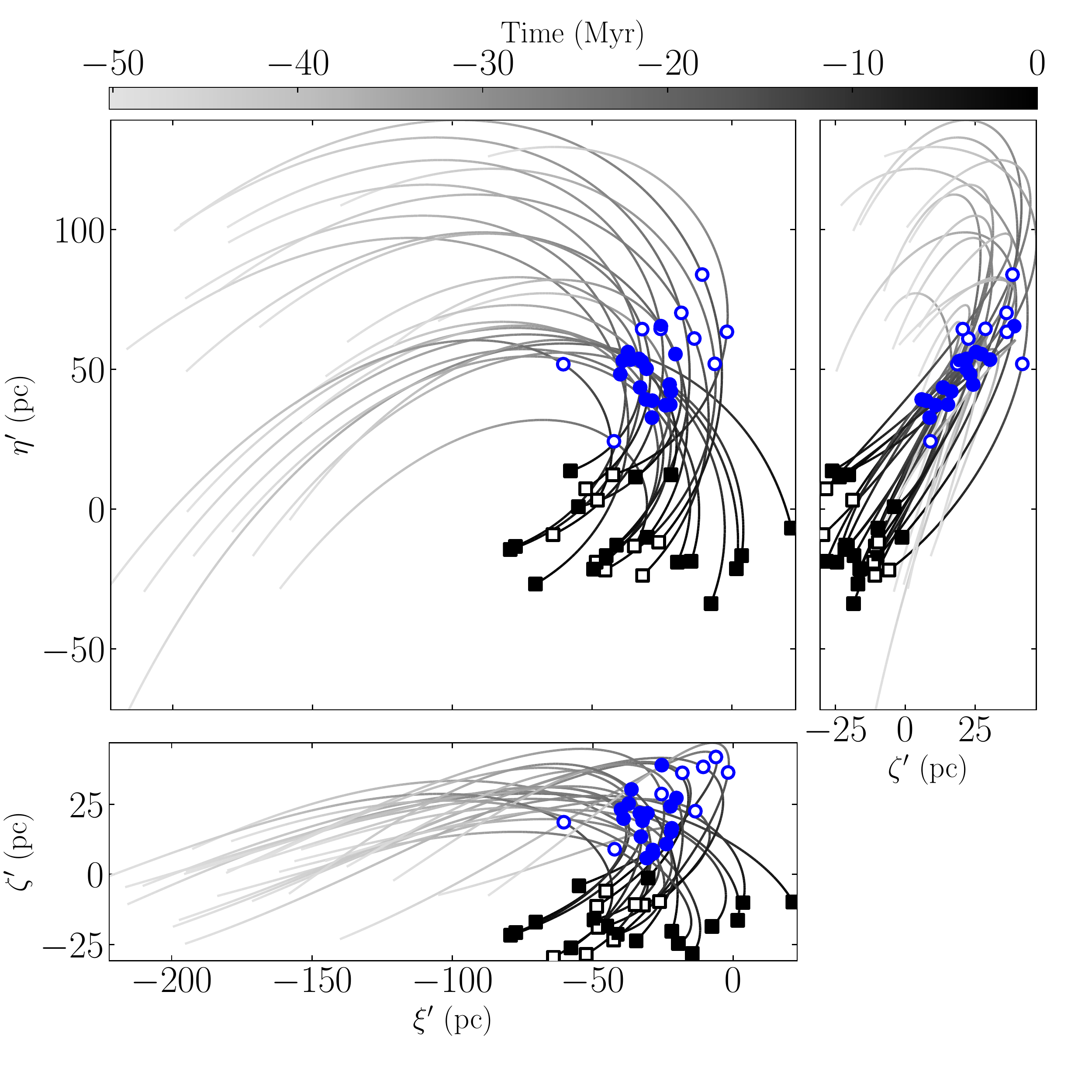}
    \caption{Orbital projection in the Galactic plane (top left) and in the two vertical planes (bottom left and top right) of our bona fide sample of 26 members of $\beta$~Pic, integrated back in time 50~Myr, under the new A\&S potential. The orbits are colour-coded with the backwards time, the black squares represent the positions in the present ($t=0$~Myr), and the blue dots represent positions at birth time ($t=-18.5$~Myr). The filled markers correspond to the core of $\beta$~Pic defined in Sect.~\ref{subsec:core}.}
    \label{fig:orbits}
\end{figure}{}

We consider the same 3D Milky Way potential as in \citet{Miret-Roig+18} to integrate the equations of motion. This model is based on the \citet{Allen+91} potential which consists of a spherical central bulge, a disc, and a massive spherical halo, but with updated parameters taken from \citet[their Table 1]{Irrgang+13}. Hereafter we refer to this model as new A\&S and we compare it with other axisymmetric models in Section~\ref{subsec:Gal-pot}. In Figure~\ref{fig:orbits}, we show the 2D orbital projections in the Galactic plane and the two vertical planes of the 26 bona fide members in our sample. The orbits have been integrated back in time 50~Myr.

\begin{figure*}
    \includegraphics[width = \textwidth]{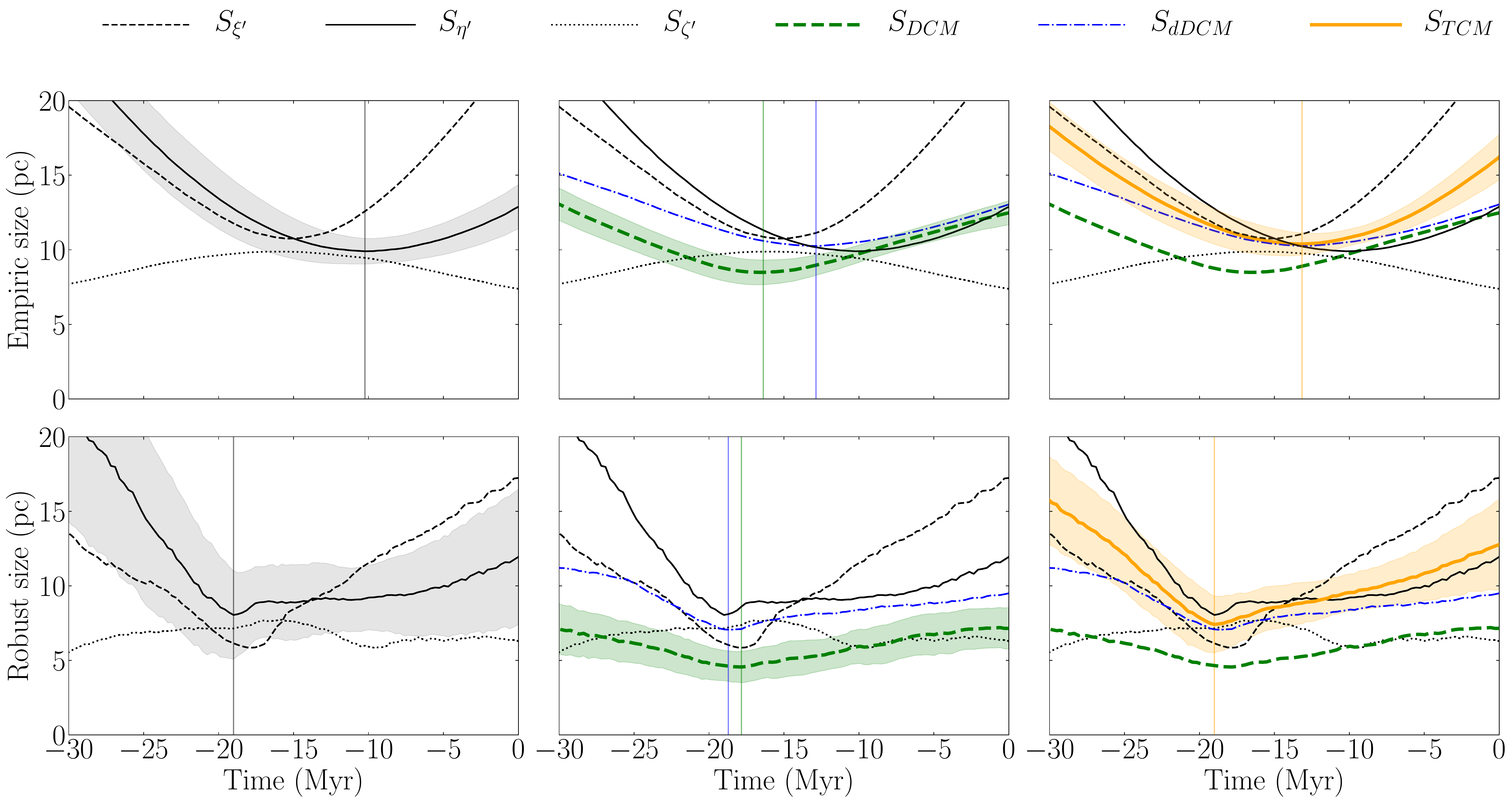}
    \caption{Size of the $\beta$~Pic association as a function of backwards time computed with the empirical covariance estimate (top panels) and the robust covariance estimate (bottom panels). The association size estimates considered in this study are indicated in the legend and described in the text.  
    The lines represent the median of 1\,000 bootstrap repetitions and the shaded areas represent the $1\sigma$ uncertainties. The orbits were integrated using the new A\&S potential. 
    }
    \label{fig:size_time}
\end{figure*}{}

\begin{figure}
    \includegraphics[width = \columnwidth]{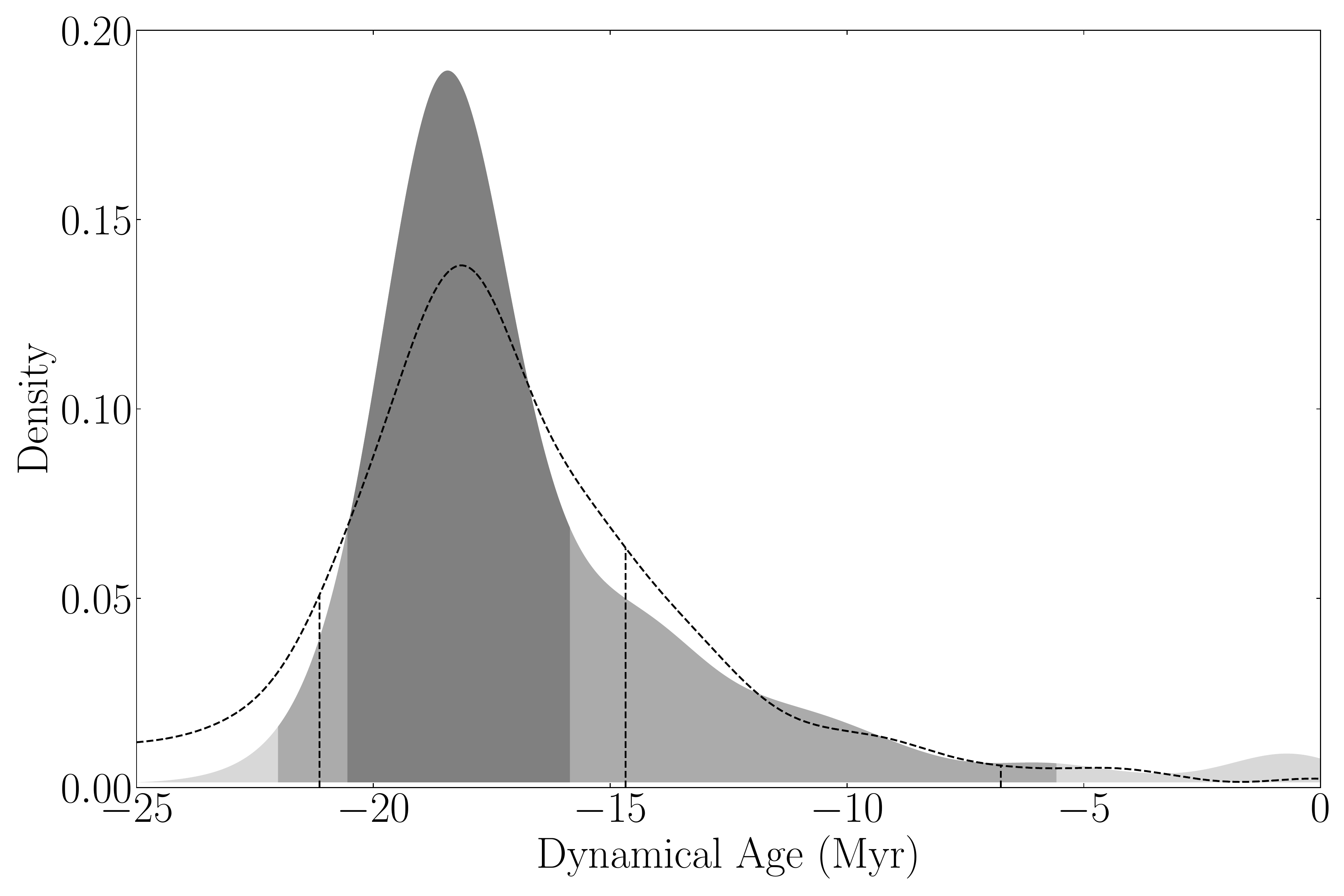}
    \caption{Dynamical age distribution of the bona fide $\beta$~Pic members, obtained with the robust estimate of the covariance matrix. The distribution obtained with the $S_{TCM}$ size estimator is colour-coded with the 68\%, 95\%, and 99.7\% highest-density intervals. The distribution obtained with the $S_{DCM}$ size estimator is shown in dashed lines and the same  highest density intervals are shown. The orbits were integrated using the new A\&S potential and we computed 1\,000 bootstrap repetitions.}
    \label{fig:hist_age}
\end{figure}{}

Following the example of  other studies \citep[e.g.][]{Fernandez08, Ducourant14, Mamajek14, Riedel17, Miret-Roig+18}, we define the dynamical age as the time at which the members of the association were most concentrated in space. The algorithm to measure the degree of concentration, hereafter the size of the association, is of uttermost importance and different strategies to compute the size have been used in the literature. These different methodologies have significantly contributed to the large spread in the dynamical traceback ages determined. In general, the size of the association is estimated with the empiric standard deviation in the spatial coordinates. However, it is very sensitive to the presence of outliers, i.e. members which significantly deviate from the mean position of the association which are not necessarily contaminants. 
In this section, we present three strategies to estimate the size of the association as a function of time. Some of them are based on classical functions used in the literature (i.e. the variance) and others are novel, representing the overall variance of the association, and independent of the coordinates chosen. In the following, we define the three functions we use to estimate the size of the association in a way that they all have units of length.

\begin{itemize}
    \item The size in the radial, azimuthal, and vertical directions ($S_{\xi^\prime}$, $S_{\eta^\prime}$, $S_{\zeta^\prime}$) are the squared root of the diagonal terms of the covariance matrix in each direction.
    \item The Trace Covariance Matrix Size ($S_{TCM}$) is defined as: 
    \begin{equation}
        S_{TCM}  =  \left[ \frac{ \Tr (\mathbf{\Sigma})}{3} \right]^{1/2}.
        \label{eq:TCM}
    \end{equation}
    \item The Determinant Covariance Matrix Size ($S_{DCM}$) is defined as:
    \begin{equation}
        S_{DCM} =  \left[ \det (\mathbf{\Sigma}) \right]^{1/6}. 
        \label{eq:DCM}
    \end{equation}
\end{itemize}

Each of these expressions are computed from the covariance matrix of the association in the configuration space. We used two different algorithms to estimate the covariance matrix, namely the empirical covariance estimation, and the robust covariance estimation, both from the Sklearn packages \citep{scikit-learn}. Whereas the first corresponds to the classical maximum likelihood estimator, the second is less sensitive to outliers in the dataset.

The size estimators $S_{\xi^\prime}$, $S_{\eta^\prime}$, and $S_{\zeta^\prime}$, when computed with the empirical covariance estimation, correspond to the classical standard deviation in each direction. 
The other two size estimators ($S_{TCM}$ and $S_{DCM}$) can be interpreted from the eigenvalues of the covariance matrix. 
The trace of the association, often referred as the total variance of the covariance matrix, coincides with the sum of its eigenvalues. In Eq.~\ref{eq:TCM}, we introduce a factor of $1/3$ (in a 3D space) so that we can interpret the $S_{TCM}$ estimator as the arithmetic mean of the variances in the individual components. In any case, this multiplicative factor changes the absolute value of the size estimator but not the locus of the minimum, which is our main interest. 
The determinant of the covariance matrix, also known as the generalised variance, can be interpreted as the geometric mean of the eigenvalues of the covariance matrix. Then, the volume of the association is proportional to the squared root of the determinant of the covariance matrix.
Finally, we define the diagonal of the Determinant Covariance Matrix Size ($S_{dDCM}$) analogously to the $S_{DCM}$ size but only considering the diagonal terms, that is, neglecting the correlations among the three spatial components. This is not a good estimator of the size of the association since it neglects part of the information included in the covariance matrix. However, it can be understood as a geometric mean of the size estimators $S_{\xi^\prime}$, $S_{\eta^\prime}$, $S_{\zeta^\prime}$, so we include it only for comparison.

In Figure~\ref{fig:size_time} we show the six parameters defining the size of $\beta$~Pic ($S_{\xi^\prime}$, $S_{\eta^\prime}$, $S_{\zeta^\prime}$, $S_{DCM}$, $S_{dDCM}$, $S_{TCM}$) computed with the empirical covariance estimate and the robust covariance estimate as a function of time. It is remarkable that the minimum size obtained with the empirical covariance estimate (top panels) depends on the size estimator, whereas we find a minimum at a similar times for all the size estimators considered with the robust covariance estimate (bottom panels). This is because the robust covariance estimate gives less weight to sources with a large dispersion, attenuating the impact of outliers. 

Going forward, we only considered the size estimates computed with the robust covariance estimates. In the left bottom panel of Fig.~\ref{fig:size_time}, we show the dispersion in the radial, azimuthal, and vertical direction, independently. We see that the vertical component does not provide useful information for constraining the age of the association, while the two components in the Galactic plane have a minimum at a similar time. In this panel we highlighted the azimuthal component ($S_{\eta^\prime}$) which is the size estimator we used in \citet{Miret-Roig+18}. 

In the middle bottom panel, we add the size from the determinant of the covariance matrix ($S_{DCM}$) and, for comparison, the inaccurate size using only the diagonal values of this matrix ($S_{dDCM}$), that is, with and without correlations, respectively. Both curves have close minima with a time difference of $\sim1$~Myr, and are also similar to the age obtained with the $S_{\eta^\prime}$ size estimator. The correlations reduce the value of the determinant and in consequence, the absolute value of $S_{DCM}$, estimating a birth size of the association of $\sim5$~pc. In the right bottom panel, we include the size estimator from the trace ($S_{TCM}$). As mentioned before, the $S_{TCM}$ and $S_{DCM}$ sizes correspond to the arithmetic and geometric mean of the eigenvalues of the covariance matrix, respectively. These two statistics are related by an inequality in which the arithmetic mean is always larger than the geometric mean and they are only equal if all the individual values are the same. This corresponds to an isotropic covariance matrix, which is not the case in our study.

Currently, thanks to the excellent astrometric precision of \textit{Gaia} and the homogeneous precise radial velocity sample derived in this work, the observational uncertainties are no longer what dominates the uncertainties in the dynamical age. We propagated the present uncertainties with an analytic approximation \citep{Miret-Roig+18} and estimated that the dispersion due to observational uncertainties is $\lesssim2$~pc at the time of minimum size. At birth, the association had a $S_{TCM}$ size of $\sim7$~pc (see Fig.~\ref{fig:size_time}), which is similar to what has been observed in star-forming regions such as Ophiuchus \citep{Canovas+2019}, Taurus \citep{Galli+2019}, and Corona Australis \citep{Galli+20}. 

As mentioned, the sample selection (i.e. the presence of contaminants or unidentified binaries) is extremely important. To estimate the impact of the sample selection on the age, we took 1\,000 random samples of the 26 bona fide $\beta$~Pic members and estimated the dynamical age with each. Then, we can determine a dynamical age and a robust uncertainty from the distribution of ages. In Figure~\ref{fig:hist_age}, we report a kernel density estimate of the age distribution with a bandwidth of 1~Myr; this value is smaller and of the order of the age uncertainties. In Table~\ref{tab:DA_diff_pots}, we report the mode and the 68\%, 95\%, and 99.7\% highest density intervals\footnote{The highest density interval is defined such that all points within the interval have a higher probability density than all points outside the interval. We used the ArviZ python package \citep{arviz_2019} to compute it. } of the age distribution. Considering the $S_{TCM}$ size estimator and the 68\% highest density interval, we find a dynamical age of $\beta$~Pic of $18.5_{-2.4}^{+2.0}$~Myr (see Table~\ref{tab:DA_diff_pots}). With the $S_{DCM}$ size estimator, we obtain a similar age, $17.6_{-2.9}^{+3.5}$~Myr. We note that the two values agree within a 1~Myr difference which is significantly smaller than the age uncertainty.

\subsection{Signs of substructure at birth time}
\label{subsec:core}

When we look at Fig.~\ref{fig:orbits}, we see that at birth, some stars appeared to be more concentrated and forming a core (filled dots), while a few members appeared to be more dispersed (empty dots). To identify these two populations, we compute the Mahalanobis distance (Eq.~\ref{eq:mahalanobis_dist}) with the robust central location and covariance of the 3D spatial distribution ($\xi^\prime, \eta^\prime, \zeta^\prime$). In Figure~\ref{fig:mahalanobis_dist_core}, we show the distribution of the Mahalanobis distances. We used the percentile $p_{68}$ to separate the core from the peripheral stars which results in 17 core stars and 9 peripheral stars (see Table~\ref{tab:26_members}). These stars were selected at birth time in the space of positions and in this space they appear most concentrated (see Fig.~\ref{fig:orbits}). 
Interestingly, in the present, the stars forming the core appear more dispersed than those originally more dispersed. In the velocity space, both populations are mixed in the present and at birth time.

It is worth mentioning that if we use only the 17 core stars to study the dynamical age, we obtain an age very similar to the value we obtained in Sect.~\ref{subsec:DA}. With the $S_{TCM}$ size we find an age of $18.8_{-2.1}^{+1.7}$~Myr and with the $S_{DCM}$ size an age of  $17.6_{-1.2}^{+3.5}$~Myr. As expected, in this case were all the stars are well concentrated at birth time, the age is independent of the covariance estimate used (empirical or robust).  
Additionally, the small bump we observe in Figure~\ref{fig:diff_pot} at $\sim-15$~Myr disappears with the age distribution obtained only with the 17 core stars. In short, if we use the core sample of 17 stars to trace back the age of $\beta$~Pic we find variations of less than 1~Myr with respect to the value we obtained in Sect.~\ref{subsec:DA}, with all the covariance and size estimates considered in this study. This is the first time that the spatial distribution of $\beta$~Pic is analysed in detail and these results should be revisited with a larger sample of members.

\subsection{Effect of the Galactic potential}\label{subsec:Gal-pot} 

\begin{figure}
    \includegraphics[width = \columnwidth]{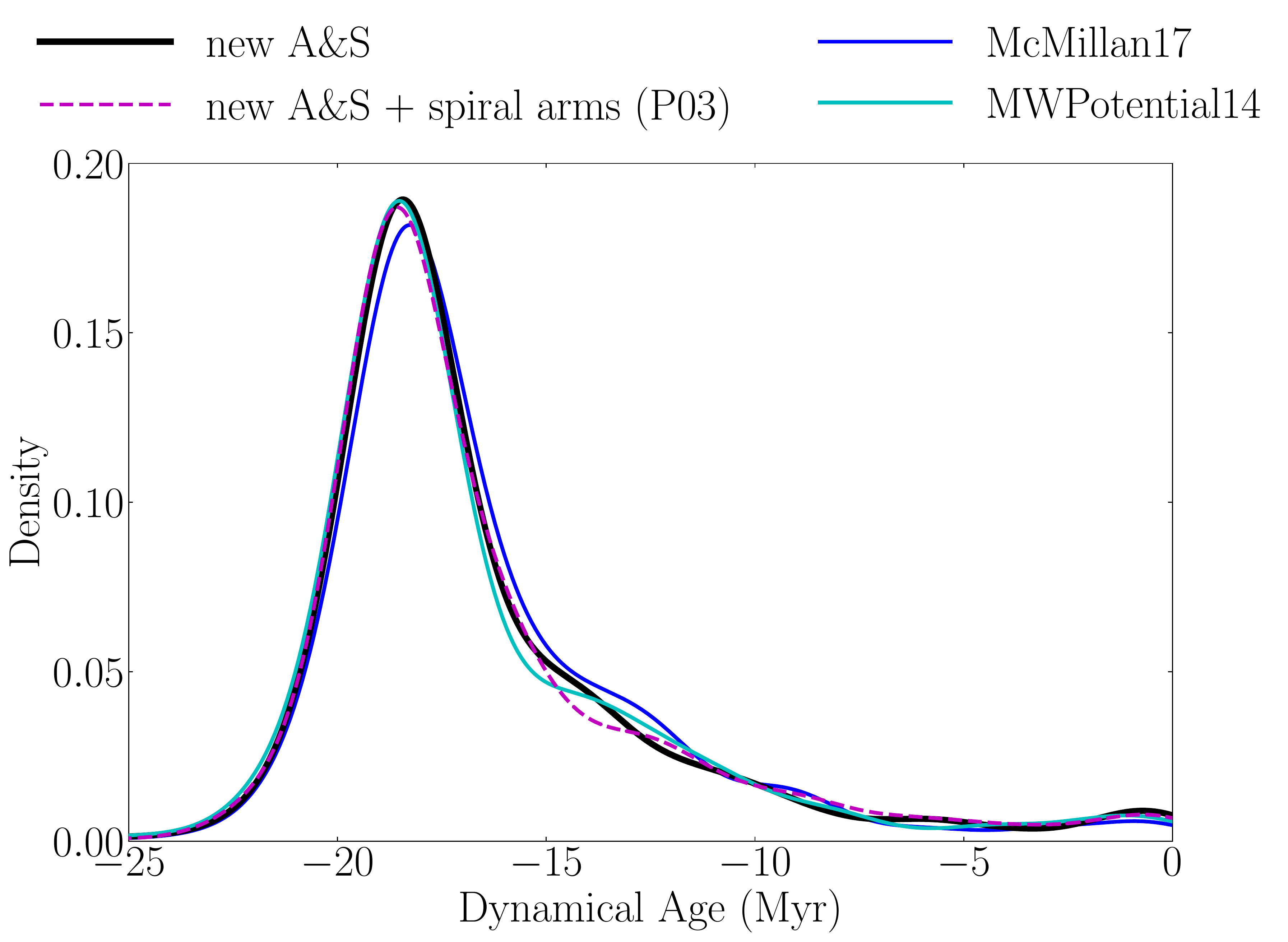}
    \caption{Dynamical age distribution of $\beta$~Pic obtained with the $S_{TCM}$ size estimator and different axisymmetric potentials (solid lines) and with the new A\&S + spiral arms (P03) potential (dashed line).    }
    \label{fig:diff_pot}
\end{figure}{}

\begin{table*}
    \centering
    \caption{Dynamical age (in Myr) obtained with the robust metrics for different potentials. We report the mode and the highest density interval for probabilities of 68\%, 95\%, and 99.7\%.}
    \label{tab:DA_diff_pots}
\begin{tabular}{l|c|ccccccc}
\hline
\hline
 Potential & Size estimator & $p_{0.15}$ & $p_{2.5}$ & $p_{16}$ & mode & $p_{84}$ & $p_{97.5}$ & $p_{99.85}$ \\
\hline
new A\&S        & $S_{DCM}$ &   -32.0 &  -27.6 &  -21.1 &  -17.6 &  -14.7 &  -6.8 &    0 \\
McMillan17      & $S_{DCM}$ &   -43.9 &  -29.3 &  -20.2 &  -17.7 &  -14.6 &  -8.1 &    0 \\
MWPotential14   & $S_{DCM}$ &   -46.5 &  -27.8 &  -21.2 &  -17.7 &  -14.6 &  -6.1 &  -1.5 \\
\hline
new A\&S + spiral arms (P03) & $S_{DCM}$ &   -46.4 &  -29.3 &  -20.8 &  -17.9 &  -14.4 &  -7.6 &    0 \\
\hline
\hline
new A\&S        & $S_{TCM}$ &   -25.2 &  -22.0 &  -20.5 &  -18.5 &  -15.9 &  -5.6 &    0 \\
McMillan17      & $S_{TCM}$ &   -26.3 &  -22.2 &  -19.7 &  -18.2 &  -15.2 &  -8.6 &    0 \\
MWPotential14   & $S_{TCM}$ &   -26.3 &  -23.7 &  -20.7 &  -18.7 &  -16.2 &  -7.6 &    0 \\
\hline
new A\&S + spiral arms (P03) & $S_{TCM}$ &   -26.4 &  -22.3 &  -20.2 &  -18.5 &  -15.9 &  -5.6 &    0 \\
\hline
\hline
\end{tabular}
     
\end{table*}{}

In this section, we discuss the effect of considering different Galactic axisymmetric potentials and including non-axisymmetric structures such as spiral arms, on the dynamical age.
First, we considered two additional axisymmetric potentials, namely the one of \citet[hereafter McMillan17]{McMillan17} and the one of \citet[hereafter MWPotential14]{Bovy15}. These two models together with the new A\&S model have similar rotation curves in the range of radius relevant here with only slight differences in the mass distribution as can be seen in their respective rotation curves (Figure D.1 in \citealt{Helmi+2018} for a comparison of new A\&S and McMillan2017 and Figure 8 in \citealt{Bovy15} for MWpotential14).

We also used a non-axisymmetric potential which accounts for the spiral arms in addition to the axisymmetric potential described in Sect. ~\ref{subsec:DA}. The 3D spiral model is the PERLAS spiral arms from \citet[hereafter new A\&S + spiral arms (P03)]{Pichardo03}. The locus is the one following \citet{Drimmel01} and has a pitch angle of 15.5\degr. We take a pattern speed of $\Omega_P=21$~km~s$^{-1}$~kpc$^{-1}$ and a mass of 0.04\% of the disc mass. These values are in agreement with the values proposed in \citet{Antoja11}.
Recently, \citet{Eilers+20} estimated a density contrast at the solar radius of 20\% which is similar to the amplitude of the arms used here which leads to a contrast of around 23\% \citep{Antoja11}. 

In Figure~\ref{fig:diff_pot} we present the dynamical age distribution obtained with different axisymmetric potentials and with the non-axisymmetric potential with spiral arms. In Table~\ref{tab:DA_diff_pots} we report the percentiles of the dynamical age distribution for each of the potentials considered. The variations in the dynamical age due to the Galactic potential are minimal, and they are all compatible with the value we obtained in Sect.~\ref{subsec:DA}. Therefore, we conclude that the variations in the dynamical age produced by different Galactic potentials are much lower than our main source of uncertainty, that is, the membership. This is valid for the potentials we have tested, the parameters of which are constrained by recent observations of the Milky Way and can be explained for the short integration times given the low age of the association.
Given that the different Galactic potentials considered here lead to changes in the dynamical age smaller than the current uncertainties, we decided to keep the results obtained with the new A\&S potential which has fewer parameters. 
%______________________________________________________________

\section{Discussion}\label{sec:discussion}

\begin{table*}
    \centering
    \caption{Literature age estimates for $\beta$~Pic. This table is an udated version of Table~1 from \citet{Mamajek14}.}
    \label{tab:age_review}
    \begin{tabular}{|l|c|l|}
        \hline
        \hline
        Reference               & Age                            & Method \\
        \hline
        \cite{Barrado99}        & $20\pm10$~Myr                  & CMD isochronal age (KM stars)\\
        \cite{Zuckerman+01}     & $12^{+8}_{-4}$~Myr             & H-R diagram isochronal age (KM stars)\\ % (GKM stars)+Li depletion \\
        \cite{Ortega02}         & $11.5$~Myr                     & Dynamical (Traceback) age \\              
        \cite{Song+03}          & $12$~Myr                       & Dynamical (Traceback) age \\     
        \cite{Ortega04}         & $10.8\pm0.3$~Myr               & Dynamical (Traceback) age \\
        \cite{Torres+06}        & $18$~Myr                       & Dynamical (Expansion) age \\             
        \cite{Makarov07}        & $31\pm21$~Myr                  & Dynamical (Traceback) age \\             
        \cite{Mentuch+08}       & $21\pm9$~Myr                   & Li depletion   \\               
        \cite{Macdonald+10}     & $\sim40$~Myr                   & Li depletion (magnetoconvection models) \\       
        \cite{Binks+14}         & $21\pm4$~Myr                   & Li depletion boundary   \\             
        \cite{Malo+14}          & $26\pm3$~Myr                   & Li depletion boundary   \\            
        \cite{Malo+14}          & $21.5\pm6.5$~Myr (15--28 Myr)  & H–R diagram isochronal age (KM stars)   \\ 
        \cite{Mamajek14}        & $22\pm3$~Myr                   & CMD isochronal age (FG stars)   \\     
        \cite{Mamajek14}        & $13-58$~Myr                    & Dynamical (Expansion) age   \\  
        \cite{Bell+2015}        & $24\pm3$~Myr                   & CMD isochronal age \\
        \cite{Messina+16}       & $25\pm3$~Myr                   & Li depletion boundary (rotation models) \\ 
        \cite{Miret-Roig+18}    & $13^{+7}_{-0}$~Myr             & Dynamical (Traceback) age   \\          
        \cite{Crundall+19}      & $18.3_{-1.2}^{+1.3}$~Myr       & Dynamical (Forward-modelling) age  \\ 
        \cite{Ujjwal+20}        & 19.38~Myr ($5.5-54.5$~Myr)     & CMD isochronal age \\
        \hline
        \textbf{This work} & \textbf{18.5$_{-2.4}^{+2.0}$~Myr} & \textbf{Dynamical (Traceback) age} \\
        \hline
        \hline
      
    \end{tabular}
\end{table*}{}

In the previous section, we discussed different strategies to determine the dynamical traceback age of $\beta$~Pic. All of them are compatible, with differences of $\lesssim1$~Myr, significantly smaller than the age uncertainties. Going forward, we adopt the an age of $18.5_{-2.4}^{+2.0}$~Myr, obtained for the sample of 26 bona fide members, with the $S_{TCM}$ size, and with the axisymmetric potential.
Our study provides the first traceback age which conforms with other dynamical ages recently published in the literature, such as the expansion or the forward-modelling algorithms and with ages based on evolutionary models such as the lithium depletion or the isochronal ages. In Table~\ref{tab:age_review}, we present a compilation of previous age estimates published in the literature and we see that our determination is compatible with the majority of them.
The first reliable age determination of the $\beta$~Pic star and its moving group was an isochronal age presented in \citet{Barrado99}, $20\pm10$~Myr, which is in full agreement with our current estimate. 

The earliest traceback studies of $\beta$~Pic obtained an age of 11--13~Myr (\citealt{Ortega02}, \citealt{Song+03}, \citealt{Ortega04}), which is younger that what we find here. These differences are most probably due to the large observational uncertainties of the pre-\textit{Gaia} astrometry and, thus, to the presence of a significant number of kinematic contaminants. Those authors used the maximum size between stars to determine the age of the association. We did not consider this size estimator in our study but it is clearly sensitive to the presence of outliers in the dataset. 

\citet{Miret-Roig+18} measured a dynamical age of $\beta$~Pic of $13^{+7}_{-0}$~Myr with a method that is very similar to the one presented in this work. We believe that the main differences between these two studies are 1) the precision of the 6D space phase positions, 2) the new sample selection based on a robust estimate of the 3D velocities covariance matrix, and 3) the new orbital analysis which uses an improved size estimator of the association. 

In our previous study, we used the \textit{Gaia} DR1 astrometry and a compilation of radial velocities from the literature without any treatment. Here, we use the improved precision of \textit{Gaia} DR2 and a uniform sample of radial velocities. The median uncertainty in the DR1 parallaxes and proper motions were 0.3~mas and 0.2~mas~yr$^{-1}$, respectively, compared to the values 0.05~mas and $<0.1$~mas~yr$^{-1}$ now available from the DR2. We discarded ten objects from our previous sample for being classified as binaries and five others do not have a radial velocity measurement in our work. These leaves only six objects in common between the two works ($23$\% of our new sample). The black solid line in Figure~\ref{fig:size_time} ($S_{\eta^\prime}$) corresponds to the methodology used in \citet{Miret-Roig+18} (their blue curve in Fig.~7). We see that with the empiric covariance (top panel) matrix we still find younger dynamical traceback ages. On the contrary, the size estimator $S_{\eta^\prime}$, if we use a robust estimate of the covariance matrix, we recover a similar age to the one reported in Sect.~\ref{subsec:DA}. 

Another technique used in the literature to measure a kinematic age consists of studying if the association is under expansion. \citet{Torres+06} found a linear relation between the velocity and the position in the direction of the Galactic centre, which results in an age of $\sim18$~Myr. In a similar approach, \citet{Mamajek14} found an age of 21$_{-5}^{+10}$~Myr, taking into account the positions and velocities in the Galactic plane. Both results are compatible with what we obtain here. In addition, we used our new accurate sample to estimate an expansion age of $\beta$~Pic. We fitted a line between the cartesian heliocentric positions $XYZ$ and velocities $UVW$. We find evidence of expansion in the direction towards the Galactic centre and in the direction of Galactic rotation with slopes of $\kappa_X= 0.057\pm0.006$~km~s~$^{-1}$~pc$^{-1}$ and $\kappa_Y=0.033 \pm0.008$~km~s~$^{-1}$~pc$^{-1}$, respectively. In the vertical direction, we find a slope of $\kappa_Z= -0.02\pm0.02$~km~s~$^{-1}$~pc$^{-1}$, which is slightly negative but compatible with zero. These coefficients result in an expansion age\footnote{To compute the expansion age we used the relation $\tau=\gamma^{-1}\kappa^{-1}$ , where $\gamma = 1.022\,712\,165$~s~pc~km$^{-1}$~Myr$^{-1}$.} of $17\pm2$~Myr and $29\pm4$~Myr in the radial and azimuthal directions. If we combine these measures with a weighted mean as done by \citet{Mamajek14}, we obtain an expansion age of $20\pm4$~Myr, in excellent agreement with our traceback age.

Recently, \citet{Crundall+19} provided a new tool (Chronostar) to determine a dynamical age applying the forward-modelling technique, and obtained an age of 18.3$_{-1.2}^{+1.3}$~Myr. It is interesting to see how similar the results of their study are to ours despite the different sample of members (we have 15 members in common, 25\% of their sample) and method used. These results prove that both methods are complementary. Their method allowed them to detect the $\beta$~Pic members among a large catalogue of field stars while ours provides a deeper orbital analysis allowing us to discover, for example, the existence of a central core and a more dispersed structure at birth time.

Finally, it is important to mention that the age estimates in the literature based on the Li depletion or isochronal fitting obtained values very similar to the one obtained here and, in general, with a lower dispersion than the dynamical age estimates obtained up to now (see Table~\ref{tab:age_review}). If we exclude the work of \citet{Macdonald+10} which obtained an age of $\sim40$~Myr, twice the other works, we obtain a median value of $21\pm4$~Myr which is in good agreement with the age we measured. This is an important result since our method is independent of evolutionary models and these are two very different strategies.

%______________________________________________________________

\section{Conclusions}
\label{sec:conclusions}

In this work we measured a dynamical, traceback age of the $\beta$~Pic moving group of $18.5_{-2.4}^{+2.0}$~Myr which is compatible with ages based on evolutionary models. Our age estimate is the first traceback age that reconciles the ages determined by the traceback method with other dynamical ages (expansion, forward modelling), lithium depletion ages, or isochronal ages.

The precision in the dynamical traceback age we achieved in this study is thanks to the combination of the \textit{Gaia} DR2 astrometry and the uniform radial velocity sample of single stars that we produced in this work. We measured the radial velocity of 81 candidate members of $\beta$~Pic in a uniform manner. For ten sources, our measure is the first radial velocity estimate. This is an important result of our work, allowing us to identify 15 kinematic outliers from our initial sample and two new potential spectroscopic binaries. 

Our improved algorithm to determine the age (based on our previous work, \citealt{Miret-Roig+18}) provides a more rigorous kinematic sample selection and an improved orbital analysis. We showed the importance of using a robust estimate of the covariance matrix  (instead of an empirical one) to minimise the impact of outliers (sources which deviate from the central locus of the association which are not necessarily contaminants). We explored different size estimators computed from the covariance matrix to determine the dynamical age (the standard deviation in different directions, the determinant, and the trace). All of them provide dynamical ages with differences of less than 1~Myr, meaning that they are compatible within the uncertainties, when computed from the robust covariance matrix. Our thorough orbital analysis allowed us to propose the existence of a central core of 17 stars which appeared more concentrated at birth time.

In this study, we show that different potentials (i.e. axysimmetrics and including the effect of spiral arms) lead to changes in the dynamical age that are within the current uncertainties. Nowadays, the major source of uncertainty in the dynamical, traceback age is the sample selection and the errors in the radial velocity estimates. For this reason, we stress the importance of choosing samples with accurate radial velocity data, with uncertainties comparable to the imminent eDR3 \textit{Gaia} release. This is crucial to reject kinematic contaminants and binaries and to ensure the success of a traceback analysis.

\begin{acknowledgements}
We acknowledge Ferran Torra for his advice in the proper use of \textit{Gaia} DR2 data.
This research has received funding from the European Research Council (ERC) under the European Union’s Horizon 2020 research and innovation programme (grant agreement No 682903, P.I. H. Bouy), and from the French State in the framework of the ”Investments for the future” Program, IdEx Bordeaux, reference ANR-10-IDEX-03-02. Based in part on observations made at 
Observatoire de Haute Provence (CNRS), France. Based on observations collected at the Centro Astronómico Hispano-Alemán (CAHA) at Calar Alto, operated jointly by Junta de Andalucía and Consejo Superior de Investigaciones Científicas (IAA-CSIC). Based on observations collected at the European Organisation for Astronomical Research in the Southern Hemisphere under programmes 0103.A-9009, 088.A-9007, 089.A-9007, 089.A-9013, 089.D-0709, 090.A-9010, 090.C-0200, 090.C-0815, 091.A-9012, 091.A-9013, 091.C-0216, 091.C-0595, 092.A-9002, 092.A-9009, 093.A-9006, 093.A-9029, 094.A-9002, 094.A-9012, 095.A-9029, 099.A-9005, 099.A-9029.
This research has been funded by the Spanish State Research Agency (AEI) Projects No.ESP2017-87676-C5-1-R and No. MDM-2017-0737 Unidad de Excelencia “María de Maeztu”- Centro de Astrobiología (INTA-CSIC).
This work was supported by the MINECO (Spanish Ministry of Economy) through grant RTI2018-095076-B-C21 (MINECO/FEDER, UE). 
This research has made use of the SIMBAD database, operated at CDS, Strasbourg, France.
This work made use of the Topcat software \citep{Taylor+2005}.
We acknowledge the Gaia Project Scientist Support Team and the Gaia Data Processing and Analysis Consortium (DPAC) for the python code  PyGaia\footnote{\url{https://github.com/agabrown/PyGaia}} 
The figures presented here were created using Matplotlib \citep{Hunter+2007}. TA has been funded by European Union's Horizon 2020 research and innovation programme under the Marie Sk{\l}odowska-Curie grant agreement No. 745617.
\end{acknowledgements}

\bibliographystyle{aa} 
\bibliography{mybiblio.bib}

\begin{thebibliography}{89}
\expandafter\ifx\csname natexlab\endcsname\relax\def\natexlab#1{#1}\fi

\bibitem[{{Aceituno} {et~al.}(2013){Aceituno}, {S{\'a}nchez}, {Grupp}, {Lillo},
  {Hern{\'a}n-Obispo}, {Benitez}, {Montoya}, {Thiele}, {Pedraz}, {Barrado},
  {Dreizler}, \& {Bean}}]{Aceituno+2013}
{Aceituno}, J., {S{\'a}nchez}, S.~F., {Grupp}, F., {et~al.} 2013, \aap, 552,
  A31

\bibitem[{{Allen} \& {Santillan}(1991)}]{Allen+91}
{Allen}, C. \& {Santillan}, A. 1991, \rmxaa, 22, 255

\bibitem[{{Alonso-Floriano} {et~al.}(2015){Alonso-Floriano}, {Caballero},
  {Cort{\'e}s-Contreras}, {Solano}, \& {Montes}}]{Alonso-Floriano2015}
{Alonso-Floriano}, F.~J., {Caballero}, J.~A., {Cort{\'e}s-Contreras}, M.,
  {Solano}, E., \& {Montes}, D. 2015, \aap, 583, A85

\bibitem[{{Antoja} {et~al.}(2011){Antoja}, {Figueras}, {Romero-G{\'o}mez},
  {Pichardo}, {Valenzuela}, \& {Moreno}}]{Antoja11}
{Antoja}, T., {Figueras}, F., {Romero-G{\'o}mez}, M., {et~al.} 2011, \mnras,
  418, 1423

\bibitem[{{Asiain} {et~al.}(1999){Asiain}, {Figueras}, \& {Torra}}]{Asiain99}
{Asiain}, R., {Figueras}, F., \& {Torra}, J. 1999, \aap, 350, 434

\bibitem[{{Baranne} {et~al.}(1996){Baranne}, {Queloz}, {Mayor}, {Adrianzyk},
  {Knispel}, {Kohler}, {Lacroix}, {Meunier}, {Rimbaud}, \& {Vin}}]{Baranne1996}
{Baranne}, A., {Queloz}, D., {Mayor}, M., {et~al.} 1996, \aaps, 119, 373

\bibitem[{{Barrado y Navasc{\'u}es}(2001)}]{Barrado2001}
{Barrado y Navasc{\'u}es}, D. 2001, Astronomical Society of the Pacific
  Conference Series, Vol. 244, {The {\ensuremath{\beta}} Pic Moving Group: a
  New Age Assesment}, ed. R.~{Jayawardhana} \& T.~{Greene}, 63

\bibitem[{{Barrado y Navascu{\'e}s} {et~al.}(1999){Barrado y Navascu{\'e}s},
  {Stauffer}, {Song}, \& {Caillault}}]{Barrado99}
{Barrado y Navascu{\'e}s}, D., {Stauffer}, J.~R., {Song}, I., \& {Caillault},
  J.-P. 1999, \apjl, 520, L123

\bibitem[{{Bell} {et~al.}(2015){Bell}, {Mamajek}, \& {Naylor}}]{Bell+2015}
{Bell}, C.~P.~M., {Mamajek}, E.~E., \& {Naylor}, T. 2015, \mnras, 454, 593

\bibitem[{{Binks} \& {Jeffries}(2014)}]{Binks+14}
{Binks}, A.~S. \& {Jeffries}, R.~D. 2014, \mnras, 438, L11

\bibitem[{{Binks} \& {Jeffries}(2016)}]{Binks+16}
{Binks}, A.~S. \& {Jeffries}, R.~D. 2016, \mnras, 455, 3345

\bibitem[{{Binks} {et~al.}(2015){Binks}, {Jeffries}, \& {Maxted}}]{Binks15}
{Binks}, A.~S., {Jeffries}, R.~D., \& {Maxted}, P.~F.~L. 2015, \mnras, 452, 173

\bibitem[{{Blanco-Cuaresma}(2019)}]{Blanco-Cuaresma2019}
{Blanco-Cuaresma}, S. 2019, \mnras, 486, 2075

\bibitem[{{Blanco-Cuaresma} {et~al.}(2014){Blanco-Cuaresma}, {Soubiran},
  {Heiter}, \& {Jofr{\'e}}}]{Blanco-Cuaresma2014}
{Blanco-Cuaresma}, S., {Soubiran}, C., {Heiter}, U., \& {Jofr{\'e}}, P. 2014,
  \aap, 569, A111

\bibitem[{{Bovy}(2015)}]{Bovy15}
{Bovy}, J. 2015, \apjs, 216, 29

\bibitem[{{C{\'a}novas} {et~al.}(2019){C{\'a}novas}, {Cantero}, {Cieza},
  {Bombrun}, {Lammers}, {Mer{\'\i}n}, {Mora}, {Ribas}, \&
  {Ru{\'\i}z-Rodr{\'\i}guez}}]{Canovas+2019}
{C{\'a}novas}, H., {Cantero}, C., {Cieza}, L., {et~al.} 2019, \aap, 626, A80

\bibitem[{{Cantat-Gaudin} {et~al.}(2018){Cantat-Gaudin}, {Jordi}, {Vallenari},
  {Bragaglia}, {Balaguer-N{\'u}{\~n}ez}, {Soubiran}, {Bossini}, {Moitinho},
  {Castro-Ginard}, {Krone-Martins}, {Casamiquela}, {Sordo}, \&
  {Carrera}}]{Cantat-Gaudin+18}
{Cantat-Gaudin}, T., {Jordi}, C., {Vallenari}, A., {et~al.} 2018, \aap, 618,
  A93

\bibitem[{{Chauvin} {et~al.}(2012){Chauvin}, {Lagrange}, {Beust}, {Bonnefoy},
  {Boccaletti}, {Apai}, {Allard}, {Ehrenreich}, {Girard}, {Mouillet}, \&
  {Rouan}}]{Chauvin2012}
{Chauvin}, G., {Lagrange}, A.~M., {Beust}, H., {et~al.} 2012, \aap, 542, A41

\bibitem[{{Crundall} {et~al.}(2019){Crundall}, {Ireland}, {Krumholz},
  {Federrath}, {{\v{Z}}erjal}, \& {Hansen}}]{Crundall+19}
{Crundall}, T.~D., {Ireland}, M.~J., {Krumholz}, M.~R., {et~al.} 2019, \mnras,
  489, 3625

\bibitem[{{de Zeeuw} {et~al.}(1999){de Zeeuw}, {Hoogerwerf}, {de Bruijne},
  {Brown}, \& {Blaauw}}]{deZeeuw99}
{de Zeeuw}, P.~T., {Hoogerwerf}, R., {de Bruijne}, J.~H.~J., {Brown}, A.~G.~A.,
  \& {Blaauw}, A. 1999, \aj, 117, 354

\bibitem[{{Drimmel} \& {Spergel}(2001)}]{Drimmel01}
{Drimmel}, R. \& {Spergel}, D.~N. 2001, \apj, 556, 181

\bibitem[{{Ducourant} {et~al.}(2014){Ducourant}, {Teixeira}, {Galli}, {Le
  Campion}, {Krone-Martins}, {Zuckerman}, {Chauvin}, \& {Song}}]{Ducourant14}
{Ducourant}, C., {Teixeira}, R., {Galli}, P.~A.~B., {et~al.} 2014, \aap, 563,
  A121

\bibitem[{{Eilers} {et~al.}(2020){Eilers}, {Hogg}, {Rix}, {Hunt}, {Fouvry}, \&
  {Buck}}]{Eilers+20}
{Eilers}, A.-C., {Hogg}, D.~W., {Rix}, H.-W., {et~al.} 2020, arXiv e-prints,
  arXiv:2003.01132

\bibitem[{{Elliott} {et~al.}(2014){Elliott}, {Bayo}, {Melo}, {Torres},
  {Sterzik}, \& {Quast}}]{Elliott+2014}
{Elliott}, P., {Bayo}, A., {Melo}, C.~H.~F., {et~al.} 2014, \aap, 568, A26

\bibitem[{{Fern{\'a}ndez} {et~al.}(2008){Fern{\'a}ndez}, {Figueras}, \&
  {Torra}}]{Fernandez08}
{Fern{\'a}ndez}, D., {Figueras}, F., \& {Torra}, J. 2008, \aap, 480, 735

\bibitem[{{Gagn{\'e}} \& {Faherty}(2018)}]{Gagne2018b}
{Gagn{\'e}}, J. \& {Faherty}, J.~K. 2018, \apj, 862, 138

\bibitem[{{Gagn{\'e}} {et~al.}(2015{\natexlab{a}}){Gagn{\'e}}, {Faherty},
  {Cruz}, {Lafreni{\'e}re}, {Doyon}, {Malo}, {Burgasser}, {Naud}, {Artigau},
  {Bouchard}, {Gizis}, \& {Albert}}]{Gagne2015b}
{Gagn{\'e}}, J., {Faherty}, J.~K., {Cruz}, K.~L., {et~al.} 2015{\natexlab{a}},
  \apjs, 219, 33

\bibitem[{{Gagn{\'e}} {et~al.}(2015{\natexlab{b}}){Gagn{\'e}},
  {Lafreni{\`e}re}, {Doyon}, {Malo}, \& {Artigau}}]{Gagne2015a}
{Gagn{\'e}}, J., {Lafreni{\`e}re}, D., {Doyon}, R., {Malo}, L., \& {Artigau},
  {\'E}. 2015{\natexlab{b}}, \apj, 798, 73

\bibitem[{{Gagn{\'e}} {et~al.}(2018){Gagn{\'e}}, {Roy-Loubier}, {Faherty},
  {Doyon}, \& {Malo}}]{Gagne2018a}
{Gagn{\'e}}, J., {Roy-Loubier}, O., {Faherty}, J.~K., {Doyon}, R., \& {Malo},
  L. 2018, \apj, 860, 43

\bibitem[{{Gaia Collaboration} {et~al.}(2018{\natexlab{a}}){Gaia
  Collaboration}, {Brown}, {Vallenari}, {Prusti}, {de Bruijne}, {Babusiaux},
  {Bailer-Jones}, {Biermann}, {Evans}, {Eyer}, {Jansen}, {Jordi}, {Klioner},
  {Lammers}, {Lindegren}, {Luri}, {Mignard}, {Panem}, {Pourbaix}, {Randich},
  {Sartoretti}, {Siddiqui}, {Soubiran}, {van Leeuwen}, {Walton}, {Arenou},
  {Bastian}, {Cropper}, {Drimmel}, {Katz}, {Lattanzi}, {Bakker}, {Cacciari},
  {Casta{\~n}eda}, {Chaoul}, {Cheek}, {De Angeli}, {Fabricius}, {Guerra},
  {Holl}, {Masana}, {Messineo}, {Mowlavi}, {Nienartowicz}, {Panuzzo},
  {Portell}, {Riello}, {Seabroke}, {Tanga}, {Th{\'e}venin}, {Gracia-Abril},
  {Comoretto}, {Garcia-Reinaldos}, {Teyssier}, {Altmann}, {Andrae}, {Audard},
  {Bellas-Velidis}, {Benson}, {Berthier}, {Blomme}, {Burgess}, {Busso},
  {Carry}, {Cellino}, {Clementini}, {Clotet}, {Creevey}, {Davidson}, {De
  Ridder}, {Delchambre}, {Dell'Oro}, {Ducourant},
  {Fern{\'a}ndez-Hern{\'a}ndez}, {Fouesneau}, {Fr{\'e}mat}, {Galluccio},
  {Garc{\'\i}a-Torres}, {Gonz{\'a}lez-N{\'u}{\~n}ez}, {Gonz{\'a}lez-Vidal},
  {Gosset}, {Guy}, {Halbwachs}, {Hambly}, {Harrison}, {Hern{\'a}ndez},
  {Hestroffer}, {Hodgkin}, {Hutton}, {Jasniewicz}, {Jean-Antoine-Piccolo},
  {Jordan}, {Korn}, {Krone-Martins}, {Lanzafame}, {Lebzelter}, {L{\"o}ffler},
  {Manteiga}, {Marrese}, {Mart{\'\i}n-Fleitas}, {Moitinho}, {Mora}, {Muinonen},
  {Osinde}, {Pancino}, {Pauwels}, {Petit}, {Recio-Blanco}, {Richards},
  {Rimoldini}, {Robin}, {Sarro}, {Siopis}, {Smith}, {Sozzetti}, {S{\"u}veges},
  {Torra}, {van Reeven}, {Abbas}, {Abreu Aramburu}, {Accart}, {Aerts},
  {Altavilla}, {{\'A}lvarez}, {Alvarez}, {Alves}, {Anderson}, {Andrei},
  {Anglada Varela}, {Antiche}, {Antoja}, {Arcay}, {Astraatmadja}, {Bach},
  {Baker}, {Balaguer-N{\'u}{\~n}ez}, {Balm}, {Barache}, {Barata}, {Barbato},
  {Barblan}, {Barklem}, {Barrado}, {Barros}, {Barstow}, {Bartholom{\'e}
  Mu{\~n}oz}, {Bassilana}, {Becciani}, {Bellazzini}, {Berihuete}, {Bertone},
  {Bianchi}, {Bienaym{\'e}}, {Blanco-Cuaresma}, {Boch}, {Boeche}, {Bombrun},
  {Borrachero}, {Bossini}, {Bouquillon}, {Bourda}, {Bragaglia}, {Bramante},
  {Breddels}, {Bressan}, {Brouillet}, {Br{\"u}semeister}, {Brugaletta},
  {Bucciarelli}, {Burlacu}, {Busonero}, {Butkevich}, {Buzzi}, {Caffau},
  {Cancelliere}, {Cannizzaro}, {Cantat-Gaudin}, {Carballo}, {Carlucci},
  {Carrasco}, {Casamiquela}, {Castellani}, {Castro-Ginard}, {Charlot},
  {Chemin}, {Chiavassa}, {Cocozza}, {Costigan}, {Cowell}, {Crifo}, {Crosta},
  {Crowley}, {Cuypers}, {Dafonte}, {Damerdji}, {Dapergolas}, {David}, {David},
  {de Laverny}, {De Luise}, {De March}, {de Martino}, {de Souza}, {de Torres},
  {Debosscher}, {del Pozo}, {Delbo}, {Delgado}, {Delgado}, {Di Matteo},
  {Diakite}, {Diener}, {Distefano}, {Dolding}, {Drazinos}, {Dur{\'a}n},
  {Edvardsson}, {Enke}, {Eriksson}, {Esquej}, {Eynard Bontemps}, {Fabre},
  {Fabrizio}, {Faigler}, {Falc{\~a}o}, {Farr{\`a}s Casas}, {Federici},
  {Fedorets}, {Fernique}, {Figueras}, {Filippi}, {Findeisen}, {Fonti},
  {Fraile}, {Fraser}, {Fr{\'e}zouls}, {Gai}, {Galleti}, {Garabato},
  {Garc{\'\i}a-Sedano}, {Garofalo}, {Garralda}, {Gavel}, {Gavras}, {Gerssen},
  {Geyer}, {Giacobbe}, {Gilmore}, {Girona}, {Giuffrida}, {Glass}, {Gomes},
  {Granvik}, {Gueguen}, {Guerrier}, {Guiraud}, {Guti{\'e}rrez-S{\'a}nchez},
  {Haigron}, {Hatzidimitriou}, {Hauser}, {Haywood}, {Heiter}, {Helmi}, {Heu},
  {Hilger}, {Hobbs}, {Hofmann}, {Holland}, {Huckle}, {Hypki}, {Icardi},
  {Jan{\ss}en}, {Jevardat de Fombelle}, {Jonker}, {Juh{\'a}sz}, {Julbe},
  {Karampelas}, {Kewley}, {Klar}, {Kochoska}, {Kohley}, {Kolenberg},
  {Kontizas}, {Kontizas}, {Koposov}, {Kordopatis}, {Kostrzewa-Rutkowska},
  {Koubsky}, {Lambert}, {Lanza}, {Lasne}, {Lavigne}, {Le Fustec}, {Le
  Poncin-Lafitte}, {Lebreton}, {Leccia}, {Leclerc}, {Lecoeur-Taibi},
  {Lenhardt}, {Leroux}, {Liao}, {Licata}, {Lindstr{\o}m}, {Lister}, {Livanou},
  {Lobel}, {L{\'o}pez}, {Managau}, {Mann}, {Mantelet}, {Marchal}, {Marchant},
  {Marconi}, {Marinoni}, {Marschalk{\'o}}, {Marshall}, {Martino}, {Marton},
  {Mary}, {Massari}, {Matijevi{\v{c}}}, {Mazeh}, {McMillan}, {Messina},
  {Michalik}, {Millar}, {Molina}, {Molinaro}, {Moln{\'a}r}, {Montegriffo},
  {Mor}, {Morbidelli}, {Morel}, {Morris}, {Mulone}, {Muraveva}, {Musella},
  {Nelemans}, {Nicastro}, {Noval}, {O'Mullane}, {Ord{\'e}novic},
  {Ord{\'o}{\~n}ez-Blanco}, {Osborne}, {Pagani}, {Pagano}, {Pailler},
  {Palacin}, {Palaversa}, {Panahi}, {Pawlak}, {Piersimoni}, {Pineau}, {Plachy},
  {Plum}, {Poggio}, {Poujoulet}, {Pr{\v{s}}a}, {Pulone}, {Racero}, {Ragaini},
  {Rambaux}, {Ramos-Lerate}, {Regibo}, {Reyl{\'e}}, {Riclet}, {Ripepi}, {Riva},
  {Rivard}, {Rixon}, {Roegiers}, {Roelens}, {Romero-G{\'o}mez}, {Rowell},
  {Royer}, {Ruiz-Dern}, {Sadowski}, {Sagrist{\`a} Sell{\'e}s}, {Sahlmann},
  {Salgado}, {Salguero}, {Sanna}, {Santana-Ros}, {Sarasso}, {Savietto},
  {Schultheis}, {Sciacca}, {Segol}, {Segovia}, {S{\'e}gransan}, {Shih},
  {Siltala}, {Silva}, {Smart}, {Smith}, {Solano}, {Solitro}, {Sordo}, {Soria
  Nieto}, {Souchay}, {Spagna}, {Spoto}, {Stampa}, {Steele},
  {Steidelm{\"u}ller}, {Stephenson}, {Stoev}, {Suess}, {Surdej}, {Szabados},
  {Szegedi-Elek}, {Tapiador}, {Taris}, {Tauran}, {Taylor}, {Teixeira},
  {Terrett}, {Teyssand ier}, {Thuillot}, {Titarenko}, {Torra Clotet}, {Turon},
  {Ulla}, {Utrilla}, {Uzzi}, {Vaillant}, {Valentini}, {Valette}, {van Elteren},
  {Van Hemelryck}, {van Leeuwen}, {Vaschetto}, {Vecchiato}, {Veljanoski},
  {Viala}, {Vicente}, {Vogt}, {von Essen}, {Voss}, {Votruba}, {Voutsinas},
  {Walmsley}, {Weiler}, {Wertz}, {Wevers}, {Wyrzykowski}, {Yoldas},
  {{\v{Z}}erjal}, {Ziaeepour}, {Zorec}, {Zschocke}, {Zucker}, {Zurbach}, \&
  {Zwitter}}]{GaiaDR2}
{Gaia Collaboration}, {Brown}, A.~G.~A., {Vallenari}, A., {et~al.}
  2018{\natexlab{a}}, \aap, 616, A1

\bibitem[{{Gaia Collaboration} {et~al.}(2018{\natexlab{b}}){Gaia
  Collaboration}, {Helmi}, {van Leeuwen}, {McMillan}, {Massari}, {Antoja},
  {Robin}, {Lindegren}, {Bastian}, {Arenou}, {Babusiaux}, {Biermann},
  {Breddels}, {Hobbs}, {Jordi}, {Pancino}, {Reyl{\'e}}, {Veljanoski}, {Brown},
  {Vallenari}, {Prusti}, {de Bruijne}, {Bailer-Jones}, {Evans}, {Eyer},
  {Jansen}, {Klioner}, {Lammers}, {Luri}, {Mignard}, {Panem}, {Pourbaix},
  {Randich}, {Sartoretti}, {Siddiqui}, {Soubiran}, {Walton}, {Cropper},
  {Drimmel}, {Katz}, {Lattanzi}, {Bakker}, {Cacciari}, {Casta{\~n}eda},
  {Chaoul}, {Cheek}, {De Angeli}, {Fabricius}, {Guerra}, {Holl}, {Masana},
  {Messineo}, {Mowlavi}, {Nienartowicz}, {Panuzzo}, {Portell}, {Riello},
  {Seabroke}, {Tanga}, {Th{\'e}venin}, {Gracia-Abril}, {Comoretto},
  {Garcia-Reinaldos}, {Teyssier}, {Altmann}, {Andrae}, {Audard},
  {Bellas-Velidis}, {Benson}, {Berthier}, {Blomme}, {Burgess}, {Busso},
  {Carry}, {Cellino}, {Clementini}, {Clotet}, {Creevey}, {Davidson}, {De
  Ridder}, {Delchambre}, {Dell'Oro}, {Ducourant},
  {Fern{\'a}ndez-Hern{\'a}ndez}, {Fouesneau}, {Fr{\'e}mat}, {Galluccio},
  {Garc{\'\i}a-Torres}, {Gonz{\'a}lez-N{\'u}{\~n}ez}, {Gonz{\'a}lez-Vidal},
  {Gosset}, {Guy}, {Halbwachs}, {Hambly}, {Harrison}, {Hern{\'a}ndez},
  {Hestroffer}, {Hodgkin}, {Hutton}, {Jasniewicz}, {Jean-Antoine-Piccolo},
  {Jordan}, {Korn}, {Krone-Martins}, {Lanzafame}, {Lebzelter}, {L{\"o}ffler},
  {Manteiga}, {Marrese}, {Mart{\'\i}n-Fleitas}, {Moitinho}, {Mora}, {Muinonen},
  {Osinde}, {Pauwels}, {Petit}, {Recio-Blanco}, {Richards}, {Rimoldini},
  {Sarro}, {Siopis}, {Smith}, {Sozzetti}, {S{\"u}veges}, {Torra}, {van Reeven},
  {Abbas}, {Abreu Aramburu}, {Accart}, {Aerts}, {Altavilla}, {{\'A}lvarez},
  {Alvarez}, {Alves}, {Anderson}, {Andrei}, {Anglada Varela}, {Antiche},
  {Arcay}, {Astraatmadja}, {Bach}, {Baker}, {Balaguer-N{\'u}{\~n}ez}, {Balm},
  {Barache}, {Barata}, {Barbato}, {Barblan}, {Barklem}, {Barrado}, {Barros},
  {Barstow}, {Bartholom{\'e} Mu{\~n}oz}, {Bassilana}, {Becciani}, {Bellazzini},
  {Berihuete}, {Bertone}, {Bianchi}, {Bienaym{\'e}}, {Blanco-Cuaresma}, {Boch},
  {Boeche}, {Bombrun}, {Borrachero}, {Bossini}, {Bouquillon}, {Bourda},
  {Bragaglia}, {Bramante}, {Bressan}, {Brouillet}, {Br{\"u}semeister},
  {Brugaletta}, {Bucciarelli}, {Burlacu}, {Busonero}, {Butkevich}, {Buzzi},
  {Caffau}, {Cancelliere}, {Cannizzaro}, {Cantat-Gaudin}, {Carballo},
  {Carlucci}, {Carrasco}, {Casamiquela}, {Castellani}, {Castro-Ginard},
  {Charlot}, {Chemin}, {Chiavassa}, {Cocozza}, {Costigan}, {Cowell}, {Crifo},
  {Crosta}, {Crowley}, {Cuypers}, {Dafonte}, {Damerdji}, {Dapergolas}, {David},
  {David}, {de Laverny}, {De Luise}, {De March}, {de Martino}, {de Souza}, {de
  Torres}, {Debosscher}, {del Pozo}, {Delbo}, {Delgado}, {Delgado}, {Di
  Matteo}, {Diakite}, {Diener}, {Distefano}, {Dolding}, {Drazinos},
  {Dur{\'a}n}, {Edvardsson}, {Enke}, {Eriksson}, {Esquej}, {Eynard Bontemps},
  {Fabre}, {Fabrizio}, {Faigler}, {Falc{\~a}o}, {Farr{\`a}s Casas}, {Federici},
  {Fedorets}, {Fernique}, {Figueras}, {Filippi}, {Findeisen}, {Fonti},
  {Fraile}, {Fraser}, {Fr{\'e}zouls}, {Gai}, {Galleti}, {Garabato},
  {Garc{\'\i}a-Sedano}, {Garofalo}, {Garralda}, {Gavel}, {Gavras}, {Gerssen},
  {Geyer}, {Giacobbe}, {Gilmore}, {Girona}, {Giuffrida}, {Glass}, {Gomes},
  {Granvik}, {Gueguen}, {Guerrier}, {Guiraud}, {Guti{\'e}rrez-S{\'a}nchez},
  {Hofmann}, {Holland}, {Huckle}, {Hypki}, {Icardi}, {Jan{\ss}en}, {Jevardat de
  Fombelle}, {Jonker}, {Juh{\'a}sz}, {Julbe}, {Karampelas}, {Kewley}, {Klar},
  {Kochoska}, {Kohley}, {Kolenberg}, {Kontizas}, {Kontizas}, {Koposov},
  {Kordopatis}, {Kostrzewa-Rutkowska}, {Koubsky}, {Lambert}, {Lanza}, {Lasne},
  {Lavigne}, {Le Fustec}, {Le Poncin-Lafitte}, {Lebreton}, {Leccia}, {Leclerc},
  {Lecoeur-Taibi}, {Lenhardt}, {Leroux}, {Liao}, {Licata}, {Lindstr{\o}m},
  {Lister}, {Livanou}, {Lobel}, {L{\'o}pez}, {Managau}, {Mann}, {Mantelet},
  {Marchal}, {Marchant}, {Marconi}, {Marinoni}, {Marschalk{\'o}}, {Marshall},
  {Martino}, {Marton}, {Mary}, {Matijevi{\v{c}}}, {Mazeh}, {Messina},
  {Michalik}, {Millar}, {Molina}, {Molinaro}, {Moln{\'a}r}, {Montegriffo},
  {Mor}, {Morbidelli}, {Morel}, {Morris}, {Mulone}, {Muraveva}, {Musella},
  {Nelemans}, {Nicastro}, {Noval}, {O'Mullane}, {Ord{\'e}novic},
  {Ord{\'o}{\~n}ez-Blanco}, {Osborne}, {Pagani}, {Pagano}, {Pailler},
  {Palacin}, {Palaversa}, {Panahi}, {Pawlak}, {Piersimoni}, {Pineau}, {Plachy},
  {Plum}, {Poggio}, {Poujoulet}, {Pr{\v{s}}a}, {Pulone}, {Racero}, {Ragaini},
  {Rambaux}, {Ramos-Lerate}, {Regibo}, {Riclet}, {Ripepi}, {Riva}, {Rivard},
  {Rixon}, {Roegiers}, {Roelens}, {Romero-G{\'o}mez}, {Rowell}, {Royer},
  {Ruiz-Dern}, {Sadowski}, {Sagrist{\`a} Sell{\'e}s}, {Sahlmann}, {Salgado},
  {Salguero}, {Sanna}, {Santana-Ros}, {Sarasso}, {Savietto}, {Schultheis},
  {Sciacca}, {Segol}, {Segovia}, {S{\'e}gransan}, {Shih}, {Siltala}, {Silva},
  {Smart}, {Smith}, {Solano}, {Solitro}, {Sordo}, {Soria Nieto}, {Souchay},
  {Spagna}, {Spoto}, {Stampa}, {Steele}, {Steidelm{\"u}ller}, {Stephenson},
  {Stoev}, {Suess}, {Surdej}, {Szabados}, {Szegedi-Elek}, {Tapiador}, {Taris},
  {Tauran}, {Taylor}, {Teixeira}, {Terrett}, {Teyssand ier}, {Thuillot},
  {Titarenko}, {Torra Clotet}, {Turon}, {Ulla}, {Utrilla}, {Uzzi}, {Vaillant},
  {Valentini}, {Valette}, {van Elteren}, {Van Hemelryck}, {van Leeuwen},
  {Vaschetto}, {Vecchiato}, {Viala}, {Vicente}, {Vogt}, {von Essen}, {Voss},
  {Votruba}, {Voutsinas}, {Walmsley}, {Weiler}, {Wertz}, {Wevems},
  {Wyrzykowski}, {Yoldas}, {{\v{Z}}erjal}, {Ziaeepour}, {Zorec}, {Zschocke},
  {Zucker}, {Zurbach}, \& {Zwitter}}]{Helmi+2018}
{Gaia Collaboration}, {Helmi}, A., {van Leeuwen}, F., {et~al.}
  2018{\natexlab{b}}, \aap, 616, A12

\bibitem[{{Gaia Collaboration} {et~al.}(2016){Gaia Collaboration}, {Prusti},
  {de Bruijne}, {Brown}, {Vallenari}, {Babusiaux}, {Bailer-Jones}, {Bastian},
  {Biermann}, {Evans}, {Eyer}, {Jansen}, {Jordi}, {Klioner}, {Lammers},
  {Lindegren}, {Luri}, {Mignard}, {Milligan}, {Panem}, {Poinsignon},
  {Pourbaix}, {Randich}, {Sarri}, {Sartoretti}, {Siddiqui}, {Soubiran},
  {Valette}, {van Leeuwen}, {Walton}, {Aerts}, {Arenou}, {Cropper}, {Drimmel},
  {H{\o}g}, {Katz}, {Lattanzi}, {O'Mullane}, {Grebel}, {Holland}, {Huc},
  {Passot}, {Bramante}, {Cacciari}, {Casta{\~n}eda}, {Chaoul}, {Cheek}, {De
  Angeli}, {Fabricius}, {Guerra}, {Hern{\'a}ndez}, {Jean-Antoine-Piccolo},
  {Masana}, {Messineo}, {Mowlavi}, {Nienartowicz}, {Ord{\'o}{\~n}ez-Blanco},
  {Panuzzo}, {Portell}, {Richards}, {Riello}, {Seabroke}, {Tanga},
  {Th{\'e}venin}, {Torra}, {Els}, {Gracia-Abril}, {Comoretto},
  {Garcia-Reinaldos}, {Lock}, {Mercier}, {Altmann}, {Andrae}, {Astraatmadja},
  {Bellas-Velidis}, {Benson}, {Berthier}, {Blomme}, {Busso}, {Carry},
  {Cellino}, {Clementini}, {Cowell}, {Creevey}, {Cuypers}, {Davidson}, {De
  Ridder}, {de Torres}, {Delchambre}, {Dell'Oro}, {Ducourant}, {Fr{\'e}mat},
  {Garc{\'\i}a-Torres}, {Gosset}, {Halbwachs}, {Hambly}, {Harrison}, {Hauser},
  {Hestroffer}, {Hodgkin}, {Huckle}, {Hutton}, {Jasniewicz}, {Jordan},
  {Kontizas}, {Korn}, {Lanzafame}, {Manteiga}, {Moitinho}, {Muinonen},
  {Osinde}, {Pancino}, {Pauwels}, {Petit}, {Recio-Blanco}, {Robin}, {Sarro},
  {Siopis}, {Smith}, {Smith}, {Sozzetti}, {Thuillot}, {van Reeven}, {Viala},
  {Abbas}, {Abreu Aramburu}, {Accart}, {Aguado}, {Allan}, {Allasia},
  {Altavilla}, {{\'A}lvarez}, {Alves}, {Anderson}, {Andrei}, {Anglada Varela},
  {Antiche}, {Antoja}, {Ant{\'o}n}, {Arcay}, {Atzei}, {Ayache}, {Bach},
  {Baker}, {Balaguer-N{\'u}{\~n}ez}, {Barache}, {Barata}, {Barbier}, {Barblan},
  {Baroni}, {Barrado y Navascu{\'e}s}, {Barros}, {Barstow}, {Becciani},
  {Bellazzini}, {Bellei}, {Bello Garc{\'\i}a}, {Belokurov}, {Bendjoya},
  {Berihuete}, {Bianchi}, {Bienaym{\'e}}, {Billebaud}, {Blagorodnova},
  {Blanco-Cuaresma}, {Boch}, {Bombrun}, {Borrachero}, {Bouquillon}, {Bourda},
  {Bouy}, {Bragaglia}, {Breddels}, {Brouillet}, {Br{\"u}semeister},
  {Bucciarelli}, {Budnik}, {Burgess}, {Burgon}, {Burlacu}, {Busonero}, {Buzzi},
  {Caffau}, {Cambras}, {Campbell}, {Cancelliere}, {Cantat-Gaudin}, {Carlucci},
  {Carrasco}, {Castellani}, {Charlot}, {Charnas}, {Charvet}, {Chassat},
  {Chiavassa}, {Clotet}, {Cocozza}, {Collins}, {Collins}, {Costigan}, {Crifo},
  {Cross}, {Crosta}, {Crowley}, {Dafonte}, {Damerdji}, {Dapergolas}, {David},
  {David}, {De Cat}, {de Felice}, {de Laverny}, {De Luise}, {De March}, {de
  Martino}, {de Souza}, {Debosscher}, {del Pozo}, {Delbo}, {Delgado},
  {Delgado}, {di Marco}, {Di Matteo}, {Diakite}, {Distefano}, {Dolding}, {Dos
  Anjos}, {Drazinos}, {Dur{\'a}n}, {Dzigan}, {Ecale}, {Edvardsson}, {Enke},
  {Erdmann}, {Escolar}, {Espina}, {Evans}, {Eynard Bontemps}, {Fabre},
  {Fabrizio}, {Faigler}, {Falc{\~a}o}, {Farr{\`a}s Casas}, {Faye}, {Federici},
  {Fedorets}, {Fern{\'a}ndez-Hern{\'a}ndez}, {Fernique}, {Fienga}, {Figueras},
  {Filippi}, {Findeisen}, {Fonti}, {Fouesneau}, {Fraile}, {Fraser}, {Fuchs},
  {Furnell}, {Gai}, {Galleti}, {Galluccio}, {Garabato}, {Garc{\'\i}a-Sedano},
  {Gar{\'e}}, {Garofalo}, {Garralda}, {Gavras}, {Gerssen}, {Geyer}, {Gilmore},
  {Girona}, {Giuffrida}, {Gomes}, {Gonz{\'a}lez-Marcos},
  {Gonz{\'a}lez-N{\'u}{\~n}ez}, {Gonz{\'a}lez-Vidal}, {Granvik}, {Guerrier},
  {Guillout}, {Guiraud}, {G{\'u}rpide}, {Guti{\'e}rrez-S{\'a}nchez}, {Guy},
  {Haigron}, {Hatzidimitriou}, {Haywood}, {Heiter}, {Helmi}, {Hobbs},
  {Hofmann}, {Holl}, {Holland }, {Hunt}, {Hypki}, {Icardi}, {Irwin}, {Jevardat
  de Fombelle}, {Jofr{\'e}}, {Jonker}, {Jorissen}, {Julbe}, {Karampelas},
  {Kochoska}, {Kohley}, {Kolenberg}, {Kontizas}, {Koposov}, {Kordopatis},
  {Koubsky}, {Kowalczyk}, {Krone-Martins}, {Kudryashova}, {Kull}, {Bachchan},
  {Lacoste-Seris}, {Lanza}, {Lavigne}, {Le Poncin-Lafitte}, {Lebreton},
  {Lebzelter}, {Leccia}, {Leclerc}, {Lecoeur-Taibi}, {Lemaitre}, {Lenhardt},
  {Leroux}, {Liao}, {Licata}, {Lindstr{\o}m}, {Lister}, {Livanou}, {Lobel},
  {L{\"o}ffler}, {L{\'o}pez}, {Lopez-Lozano}, {Lorenz}, {Loureiro},
  {MacDonald}, {Magalh{\~a}es Fernandes}, {Managau}, {Mann}, {Mantelet},
  {Marchal}, {Marchant}, {Marconi}, {Marie}, {Marinoni}, {Marrese},
  {Marschalk{\'o}}, {Marshall}, {Mart{\'\i}n-Fleitas}, {Martino}, {Mary},
  {Matijevi{\v{c}}}, {Mazeh}, {McMillan}, {Messina}, {Mestre}, {Michalik},
  {Millar}, {Miranda}, {Molina}, {Molinaro}, {Molinaro}, {Moln{\'a}r},
  {Moniez}, {Montegriffo}, {Monteiro}, {Mor}, {Mora}, {Morbidelli}, {Morel},
  {Morgenthaler}, {Morley}, {Morris}, {Mulone}, {Muraveva}, {Musella},
  {Narbonne}, {Nelemans}, {Nicastro}, {Noval}, {Ord{\'e}novic},
  {Ordieres-Mer{\'e}}, {Osborne}, {Pagani}, {Pagano}, {Pailler}, {Palacin},
  {Palaversa}, {Parsons}, {Paulsen}, {Pecoraro}, {Pedrosa}, {Pentik{\"a}inen},
  {Pereira}, {Pichon}, {Piersimoni}, {Pineau}, {Plachy}, {Plum}, {Poujoulet},
  {Pr{\v{s}}a}, {Pulone}, {Ragaini}, {Rago}, {Rambaux}, {Ramos-Lerate},
  {Ranalli}, {Rauw}, {Read}, {Regibo}, {Renk}, {Reyl{\'e}}, {Ribeiro},
  {Rimoldini}, {Ripepi}, {Riva}, {Rixon}, {Roelens}, {Romero-G{\'o}mez},
  {Rowell}, {Royer}, {Rudolph}, {Ruiz-Dern}, {Sadowski}, {Sagrist{\`a}
  Sell{\'e}s}, {Sahlmann}, {Salgado}, {Salguero}, {Sarasso}, {Savietto},
  {Schnorhk}, {Schultheis}, {Sciacca}, {Segol}, {Segovia}, {Segransan},
  {Serpell}, {Shih}, {Smareglia}, {Smart}, {Smith}, {Solano}, {Solitro},
  {Sordo}, {Soria Nieto}, {Souchay}, {Spagna}, {Spoto}, {Stampa}, {Steele},
  {Steidelm{\"u}ller}, {Stephenson}, {Stoev}, {Suess}, {S{\"u}veges}, {Surdej},
  {Szabados}, {Szegedi-Elek}, {Tapiador}, {Taris}, {Tauran}, {Taylor},
  {Teixeira}, {Terrett}, {Tingley}, {Trager}, {Turon}, {Ulla}, {Utrilla},
  {Valentini}, {van Elteren}, {Van Hemelryck}, {van Leeuwen}, {Varadi},
  {Vecchiato}, {Veljanoski}, {Via}, {Vicente}, {Vogt}, {Voss}, {Votruba},
  {Voutsinas}, {Walmsley}, {Weiler}, {Weingrill}, {Werner}, {Wevers},
  {Whitehead}, {Wyrzykowski}, {Yoldas}, {{\v{Z}}erjal}, {Zucker}, {Zurbach},
  {Zwitter}, {Alecu}, {Allen}, {Allende Prieto}, {Amorim},
  {Anglada-Escud{\'e}}, {Arsenijevic}, {Azaz}, {Balm}, {Beck}, {Bernstein},
  {Bigot}, {Bijaoui}, {Blasco}, {Bonfigli}, {Bono}, {Boudreault}, {Bressan},
  {Brown}, {Brunet}, {Bunclark}, {Buonanno}, {Butkevich}, {Carret}, {Carrion},
  {Chemin}, {Ch{\'e}reau}, {Corcione}, {Darmigny}, {de Boer}, {de Teodoro}, {de
  Zeeuw}, {Delle Luche}, {Domingues}, {Dubath}, {Fodor}, {Fr{\'e}zouls},
  {Fries}, {Fustes}, {Fyfe}, {Gallardo}, {Gallegos}, {Gardiol}, {Gebran},
  {Gomboc}, {G{\'o}mez}, {Grux}, {Gueguen}, {Heyrovsky}, {Hoar}, {Iannicola},
  {Isasi Parache}, {Janotto}, {Joliet}, {Jonckheere}, {Keil}, {Kim},
  {Klagyivik}, {Klar}, {Knude}, {Kochukhov}, {Kolka}, {Kos}, {Kutka}, {Lainey},
  {LeBouquin}, {Liu}, {Loreggia}, {Makarov}, {Marseille}, {Martayan},
  {Martinez-Rubi}, {Massart}, {Meynadier}, {Mignot}, {Munari}, {Nguyen},
  {Nordlander}, {Ocvirk}, {O'Flaherty}, {Olias Sanz}, {Ortiz}, {Osorio},
  {Oszkiewicz}, {Ouzounis}, {Palmer}, {Park}, {Pasquato}, {Peltzer}, {Peralta},
  {P{\'e}turaud}, {Pieniluoma}, {Pigozzi}, {Poels}, {Prat}, {Prod'homme},
  {Raison}, {Rebordao}, {Risquez}, {Rocca-Volmerange}, {Rosen}, {Ruiz-Fuertes},
  {Russo}, {Sembay}, {Serraller Vizcaino}, {Short}, {Siebert}, {Silva},
  {Sinachopoulos}, {Slezak}, {Soffel}, {Sosnowska}, {Strai{\v{z}}ys}, {ter
  Linden}, {Terrell}, {Theil}, {Tiede}, {Troisi}, {Tsalmantza}, {Tur},
  {Vaccari}, {Vachier}, {Valles}, {Van Hamme}, {Veltz}, {Virtanen}, {Wallut},
  {Wichmann}, {Wilkinson}, {Ziaeepour}, \& {Zschocke}}]{GaiaCol+16}
{Gaia Collaboration}, {Prusti}, T., {de Bruijne}, J.~H.~J., {et~al.} 2016,
  \aap, 595, A1

\bibitem[{{Galli} {et~al.}(2020){Galli}, {Bouy}, {Olivares}, {Miret-Roig},
  {Sarro}, {Barrado}, {Berihuete}, \& {Brandner}}]{Galli+20}
{Galli}, P.~A.~B., {Bouy}, H., {Olivares}, J., {et~al.} 2020, \aap, 634, A98

\bibitem[{{Galli} {et~al.}(2019){Galli}, {Loinard}, {Bouy}, {Sarro},
  {Ortiz-Le{\'o}n}, {Dzib}, {Olivares}, {Heyer}, {Hernandez},
  {Rom{\'a}n-Z{\'u}{\~n}iga}, {Kounkel}, \& {Covey}}]{Galli+2019}
{Galli}, P.~A.~B., {Loinard}, L., {Bouy}, H., {et~al.} 2019, \aap, 630, A137

\bibitem[{{Gray} {et~al.}(2006){Gray}, {Corbally}, {Garrison}, {McFadden},
  {Bubar}, {McGahee}, {O'Donoghue}, \& {Knox}}]{Gray+2006}
{Gray}, R.~O., {Corbally}, C.~J., {Garrison}, R.~F., {et~al.} 2006, \aj, 132,
  161

\bibitem[{{Hartman} {et~al.}(2011){Hartman}, {Bakos}, {Noyes}, {Sip{\H{o}}cz},
  {Kov{\'a}cs}, {Mazeh}, {Shporer}, \& {P{\'a}l}}]{Hartman+2011}
{Hartman}, J.~D., {Bakos}, G.~{\'A}., {Noyes}, R.~W., {et~al.} 2011, \aj, 141,
  166

\bibitem[{{Hennebelle} \& {Falgarone}(2012)}]{Hennebelle+2012}
{Hennebelle}, P. \& {Falgarone}, E. 2012, \aapr, 20, 55

\bibitem[{{Heyer} \& {Dame}(2015)}]{Heyer+2015}
{Heyer}, M. \& {Dame}, T.~M. 2015, \araa, 53, 583

\bibitem[{{Houk}(1982)}]{Houk1982}
{Houk}, N. 1982, {Michigan Catalogue of Two-dimensional Spectral Types for the
  HD stars. Volume\_3. Declinations $-40\degr$ to $-26\degr$.}

\bibitem[{Hunter(2007)}]{Hunter+2007}
Hunter, J.~D. 2007, Computing in Science \& Engineering, 9, 90

\bibitem[{{Irrgang} {et~al.}(2013){Irrgang}, {Wilcox}, {Tucker}, \&
  {Schiefelbein}}]{Irrgang+13}
{Irrgang}, A., {Wilcox}, B., {Tucker}, E., \& {Schiefelbein}, L. 2013, \aap,
  549, A137

\bibitem[{{Janson} {et~al.}(2017){Janson}, {Durkan}, {Hippler}, {Dai},
  {Brandner}, {Schlieder}, {Bonnefoy}, \& {Henning}}]{Janson+2017}
{Janson}, M., {Durkan}, S., {Hippler}, S., {et~al.} 2017, \aap, 599, A70

\bibitem[{{Jayawardhana}(2000)}]{Jayawardhana00}
{Jayawardhana}, R. 2000, Science, 288, 64

\bibitem[{{Kalas} \& {Jewitt}(1995)}]{Kalas1995}
{Kalas}, P. \& {Jewitt}, D. 1995, \aj, 110, 794

\bibitem[{{Kalas} {et~al.}(2004){Kalas}, {Liu}, \& {Matthews}}]{Kalas2004}
{Kalas}, P., {Liu}, M.~C., \& {Matthews}, B.~C. 2004, Science, 303, 1990

\bibitem[{{Kaufer} {et~al.}(1999){Kaufer}, {Stahl}, {Tubbesing},
  {N{\o}rregaard}, {Avila}, {Francois}, {Pasquini}, \& {Pizzella}}]{Kaufer1999}
{Kaufer}, A., {Stahl}, O., {Tubbesing}, S., {et~al.} 1999, The Messenger, 95, 8

\bibitem[{{Kharchenko} {et~al.}(2007){Kharchenko}, {Scholz}, {Piskunov},
  {R{\"o}ser}, \& {Schilbach}}]{Kharchenko2007}
{Kharchenko}, N.~V., {Scholz}, R.~D., {Piskunov}, A.~E., {R{\"o}ser}, S., \&
  {Schilbach}, E. 2007, Astronomische Nachrichten, 328, 889

\bibitem[{{Kiefer} {et~al.}(2014){Kiefer}, {Lecavelier des Etangs}, {Boissier},
  {Vidal-Madjar}, {Beust}, {Lagrange}, {H{\'e}brard}, \& {Ferlet}}]{Kiefer2014}
{Kiefer}, F., {Lecavelier des Etangs}, A., {Boissier}, J., {et~al.} 2014, \nat,
  514, 462

\bibitem[{{Klutsch} {et~al.}(2014){Klutsch}, {Freire Ferrero}, {Guillout},
  {Frasca}, {Marilli}, \& {Montes}}]{Klutsch+2014}
{Klutsch}, A., {Freire Ferrero}, R., {Guillout}, P., {et~al.} 2014, \aap, 567,
  A52

\bibitem[{{Kraus} {et~al.}(2017){Kraus}, {Herczeg}, {Rizzuto}, {Mann},
  {Slesnick}, {Carpenter}, {Hillenbrand}, \& {Mamajek}}]{Kraus+17}
{Kraus}, A.~L., {Herczeg}, G.~J., {Rizzuto}, A.~C., {et~al.} 2017, \apj, 838,
  150

\bibitem[{Kumar {et~al.}(2019)Kumar, Carroll, Hartikainen, \&
  Martin}]{arviz_2019}
Kumar, R., Carroll, C., Hartikainen, A., \& Martin, O.~A. 2019, The Journal of
  Open Source Software

\bibitem[{{Lagrange} {et~al.}(2010){Lagrange}, {Bonnefoy}, {Chauvin}, {Apai},
  {Ehrenreich}, {Boccaletti}, {Gratadour}, {Rouan}, {Mouillet}, {Lacour}, \&
  {Kasper}}]{Lagrange2010}
{Lagrange}, A.~M., {Bonnefoy}, M., {Chauvin}, G., {et~al.} 2010, Science, 329,
  57

\bibitem[{{Lagrange} {et~al.}(2019){Lagrange}, {Meunier}, {Rubini}, {Keppler},
  {Galland}, {Chapellier}, {Michel}, {Balona}, {Beust}, {Guillot}, {Grandjean},
  {Borgniet}, {M{\'e}karnia}, {Wilson}, {Kiefer}, {Bonnefoy}, {Lillo-Box},
  {Pantoja}, {Jones}, {Iglesias}, {Rodet}, {Diaz}, {Zapata}, {Abe}, \&
  {Schmider}}]{Lagrange2019}
{Lagrange}, A.~M., {Meunier}, N., {Rubini}, P., {et~al.} 2019, Nature
  Astronomy, 3, 1135

\bibitem[{{Lillo-Box} {et~al.}(2020){Lillo-Box}, {Aceituno}, {Pedraz},
  {Bergond}, {Galad{\'\i}-Enr{\'\i}quez}, {Azzaro}, {Arroyo-Torres},
  {Fern{\'a}ndez-Mart{\'\i}n}, {Guijarro}, {Hedrosa}, {Hermelo}, {Hoyo}, \&
  {Mart{\'\i}n-Fern{\'a}ndez}}]{Lillo-Box+2020}
{Lillo-Box}, J., {Aceituno}, J., {Pedraz}, S., {et~al.} 2020, \mnras, 491, 4496

\bibitem[{{Lindegren} {et~al.}(2018){Lindegren}, {Hern{\'a}ndez}, {Bombrun},
  {Klioner}, {Bastian}, {Ramos-Lerate}, {de Torres}, {Steidelm{\"u}ller},
  {Stephenson}, {Hobbs}, {Lammers}, {Biermann}, {Geyer}, {Hilger}, {Michalik},
  {Stampa}, {McMillan}, {Casta{\~n}eda}, {Clotet}, {Comoretto}, {Davidson},
  {Fabricius}, {Gracia}, {Hambly}, {Hutton}, {Mora}, {Portell}, {van Leeuwen},
  {Abbas}, {Abreu}, {Altmann}, {Andrei}, {Anglada}, {Balaguer-N{\'u}{\~n}ez},
  {Barache}, {Becciani}, {Bertone}, {Bianchi}, {Bouquillon}, {Bourda},
  {Br{\"u}semeister}, {Bucciarelli}, {Busonero}, {Buzzi}, {Cancelliere},
  {Carlucci}, {Charlot}, {Cheek}, {Crosta}, {Crowley}, {de Bruijne}, {de
  Felice}, {Drimmel}, {Esquej}, {Fienga}, {Fraile}, {Gai}, {Garralda},
  {Gonz{\'a}lez-Vidal}, {Guerra}, {Hauser}, {Hofmann}, {Holl}, {Jordan},
  {Lattanzi}, {Lenhardt}, {Liao}, {Licata}, {Lister}, {L{\"o}ffler},
  {Marchant}, {Martin-Fleitas}, {Messineo}, {Mignard}, {Morbidelli}, {Poggio},
  {Riva}, {Rowell}, {Salguero}, {Sarasso}, {Sciacca}, {Siddiqui}, {Smart},
  {Spagna}, {Steele}, {Taris}, {Torra}, {van Elteren}, {van Reeven}, \&
  {Vecchiato}}]{Lindegren+2018}
{Lindegren}, L., {Hern{\'a}ndez}, J., {Bombrun}, A., {et~al.} 2018, \aap, 616,
  A2

\bibitem[{{L{\'o}pez-Santiago} {et~al.}(2006){L{\'o}pez-Santiago}, {Montes},
  {Crespo-Chac{\'o}n}, \& {Fern{\'a}ndez-Figueroa}}]{Lopez-Santiago+06}
{L{\'o}pez-Santiago}, J., {Montes}, D., {Crespo-Chac{\'o}n}, I., \&
  {Fern{\'a}ndez-Figueroa}, M.~J. 2006, \apj, 643, 1160

\bibitem[{{Macdonald} \& {Mullan}(2010)}]{Macdonald+10}
{Macdonald}, J. \& {Mullan}, D.~J. 2010, \apj, 723, 1599

\bibitem[{{Makarov}(2007)}]{Makarov07}
{Makarov}, V.~V. 2007, \apjs, 169, 105

\bibitem[{{Malo} {et~al.}(2014{\natexlab{a}}){Malo}, {Artigau}, {Doyon},
  {Lafreni{\`e}re}, {Albert}, \& {Gagn{\'e}}}]{Malo2014}
{Malo}, L., {Artigau}, {\'E}., {Doyon}, R., {et~al.} 2014{\natexlab{a}}, \apj,
  788, 81

\bibitem[{{Malo} {et~al.}(2014{\natexlab{b}}){Malo}, {Doyon}, {Feiden},
  {Albert}, {Lafreni{\`e}re}, {Artigau}, {Gagn{\'e}}, \& {Riedel}}]{Malo+14}
{Malo}, L., {Doyon}, R., {Feiden}, G.~A., {et~al.} 2014{\natexlab{b}}, \apj,
  792, 37

\bibitem[{{Malo} {et~al.}(2013){Malo}, {Doyon}, {Lafreni{\`e}re}, {Artigau},
  {Gagn{\'e}}, {Baron}, \& {Riedel}}]{Malo2013}
{Malo}, L., {Doyon}, R., {Lafreni{\`e}re}, D., {et~al.} 2013, \apj, 762, 88

\bibitem[{{Mamajek} \& {Bell}(2014)}]{Mamajek14}
{Mamajek}, E.~E. \& {Bell}, C.~P.~M. 2014, \mnras, 445, 2169

\bibitem[{{McMillan}(2017)}]{McMillan17}
{McMillan}, P.~J. 2017, \mnras, 465, 76

\bibitem[{{Mentuch} {et~al.}(2008){Mentuch}, {Brandeker}, {van Kerkwijk},
  {Jayawardhana}, \& {Hauschildt}}]{Mentuch+08}
{Mentuch}, E., {Brandeker}, A., {van Kerkwijk}, M.~H., {Jayawardhana}, R., \&
  {Hauschildt}, P.~H. 2008, \apj, 689, 1127

\bibitem[{{Messina} {et~al.}(2016){Messina}, {Lanzafame}, {Feiden}, {Millward},
  {Desidera}, {Buccino}, {Curtis}, {Jofr{\'e}}, {Kehusmaa}, {Medhi}, {Monard},
  \& {Petrucci}}]{Messina+16}
{Messina}, S., {Lanzafame}, A.~C., {Feiden}, G.~A., {et~al.} 2016, \aap, 596,
  A29

\bibitem[{{Messina} {et~al.}(2017){Messina}, {Millward}, {Buccino}, {Zhang},
  {Medhi}, {Jofr{\'e}}, {Petrucci}, {Pi}, {Hambsch}, {Kehusmaa}, {Harlingten},
  {Artemenko}, {Curtis}, {Hentunen}, {Malo}, {Mauas}, {Monard}, {Muro Serrano},
  {Naves}, {Santallo}, {Savuskin}, \& {Tan}}]{Messina2017}
{Messina}, S., {Millward}, M., {Buccino}, A., {et~al.} 2017, \aap, 600, A83

\bibitem[{{Miret-Roig} {et~al.}(2018){Miret-Roig}, {Antoja},
  {Romero-G{\'o}mez}, \& {Figueras}}]{Miret-Roig+18}
{Miret-Roig}, N., {Antoja}, T., {Romero-G{\'o}mez}, M., \& {Figueras}, F. 2018,
  \aap, 615, A51

\bibitem[{{Ortega} {et~al.}(2002){Ortega}, {de la Reza}, {Jilinski}, \&
  {Bazzanella}}]{Ortega02}
{Ortega}, V.~G., {de la Reza}, R., {Jilinski}, E., \& {Bazzanella}, B. 2002,
  \apjl, 575, L75

\bibitem[{{Ortega} {et~al.}(2004){Ortega}, {de la Reza}, {Jilinski}, \&
  {Bazzanella}}]{Ortega04}
{Ortega}, V.~G., {de la Reza}, R., {Jilinski}, E., \& {Bazzanella}, B. 2004,
  \apj, 609, 243

\bibitem[{Pedregosa {et~al.}(2011)Pedregosa, Varoquaux, Gramfort, Michel,
  Thirion, Grisel, Blondel, Prettenhofer, Weiss, Dubourg, Vanderplas, Passos,
  Cournapeau, Brucher, Perrot, \& Duchesnay}]{scikit-learn}
Pedregosa, F., Varoquaux, G., Gramfort, A., {et~al.} 2011, Journal of Machine
  Learning Research, 12, 2825

\bibitem[{{Pepe} {et~al.}(2002){Pepe}, {Mayor}, {Galland}, {Naef}, {Queloz},
  {Santos}, {Udry}, \& {Burnet}}]{Pepe2002}
{Pepe}, F., {Mayor}, M., {Galland}, F., {et~al.} 2002, \aap, 388, 632

\bibitem[{{Perruchot} {et~al.}(2008){Perruchot}, {Kohler}, {Bouchy}, {Richaud},
  {Richaud}, {Moreaux}, {Merzougui}, {Sottile}, {Hill}, {Knispel}, {Regal},
  {Meunier}, {Ilovaisky}, {Le Coroller}, {Gillet}, {Schmitt}, {Pepe}, {Fleury},
  {Sosnowska}, {Vors}, {M{\'e}gevand}, {Blanc}, {Carol}, {Point}, {Laloge}, \&
  {Brunel}}]{Perruchot+2008}
{Perruchot}, S., {Kohler}, D., {Bouchy}, F., {et~al.} 2008, in Society of
  Photo-Optical Instrumentation Engineers (SPIE) Conference Series, Vol. 7014,
  \procspie, 70140J

\bibitem[{{Pichardo} {et~al.}(2003){Pichardo}, {Martos}, {Moreno}, \&
  {Espresate}}]{Pichardo03}
{Pichardo}, B., {Martos}, M., {Moreno}, E., \& {Espresate}, J. 2003, \apj, 582,
  230

\bibitem[{{Riaz} {et~al.}(2006){Riaz}, {Gizis}, \& {Harvin}}]{Riaz+2006}
{Riaz}, B., {Gizis}, J.~E., \& {Harvin}, J. 2006, \aj, 132, 866

\bibitem[{{Riedel} {et~al.}(2017{\natexlab{a}}){Riedel}, {Alam}, {Rice},
  {Cruz}, \& {Henry}}]{Riedel+2017b}
{Riedel}, A.~R., {Alam}, M.~K., {Rice}, E.~L., {Cruz}, K.~L., \& {Henry}, T.~J.
  2017{\natexlab{a}}, \apj, 840, 87

\bibitem[{{Riedel} {et~al.}(2017{\natexlab{b}}){Riedel}, {Blunt}, {Lambrides},
  {Rice}, {Cruz}, \& {Faherty}}]{Riedel17}
{Riedel}, A.~R., {Blunt}, S.~C., {Lambrides}, E.~L., {et~al.}
  2017{\natexlab{b}}, \aj, 153, 95

\bibitem[{{Schlieder} {et~al.}(2012){Schlieder}, {L{\'e}pine}, \&
  {Simon}}]{Schlieder2012}
{Schlieder}, J.~E., {L{\'e}pine}, S., \& {Simon}, M. 2012, \aj, 143, 80

\bibitem[{{Schneider} {et~al.}(2019){Schneider}, {Shkolnik}, {Allers}, {Kraus},
  {Liu}, {Weinberger}, \& {Flagg}}]{Schneider+19}
{Schneider}, A.~C., {Shkolnik}, E.~L., {Allers}, K.~N., {et~al.} 2019, \aj,
  157, 234

\bibitem[{{Sch{\"o}nrich} {et~al.}(2010){Sch{\"o}nrich}, {Binney}, \&
  {Dehnen}}]{Schonrich10}
{Sch{\"o}nrich}, R., {Binney}, J., \& {Dehnen}, W. 2010, \mnras, 403, 1829

\bibitem[{{Shkolnik} {et~al.}(2017){Shkolnik}, {Allers}, {Kraus}, {Liu}, \&
  {Flagg}}]{Shkolnik+17}
{Shkolnik}, E.~L., {Allers}, K.~N., {Kraus}, A.~L., {Liu}, M.~C., \& {Flagg},
  L. 2017, \aj, 154, 69

\bibitem[{{Shkolnik} {et~al.}(2012){Shkolnik}, {Anglada-Escud{\'e}}, {Liu},
  {Bowler}, {Weinberger}, {Boss}, {Reid}, \& {Tamura}}]{Shkolnik2012}
{Shkolnik}, E.~L., {Anglada-Escud{\'e}}, G., {Liu}, M.~C., {et~al.} 2012, \apj,
  758, 56

\bibitem[{{Song} {et~al.}(2003){Song}, {Zuckerman}, \& {Bessell}}]{Song+03}
{Song}, I., {Zuckerman}, B., \& {Bessell}, M.~S. 2003, \apj, 599, 342

\bibitem[{{Soubiran} {et~al.}(2018){Soubiran}, {Jasniewicz}, {Chemin},
  {Zurbach}, {Brouillet}, {Panuzzo}, {Sartoretti}, {Katz}, {Le Campion},
  {Marchal}, {Hestroffer}, {Th{\'e}venin}, {Crifo}, {Udry}, {Cropper},
  {Seabroke}, {Viala}, {Benson}, {Blomme}, {Jean-Antoine}, {Huckle}, {Smith},
  {Baker}, {Damerdji}, {Dolding}, {Fr{\'e}mat}, {Gosset}, {Guerrier}, {Guy},
  {Haigron}, {Jan{\ss}en}, {Plum}, {Fabre}, {Lasne}, {Pailler}, {Panem},
  {Riclet}, {Royer}, {Tauran}, {Zwitter}, {Gueguen}, \& {Turon}}]{Soubiran+18}
{Soubiran}, C., {Jasniewicz}, G., {Chemin}, L., {et~al.} 2018, \aap, 616, A7

\bibitem[{{Taylor}(2005)}]{Taylor+2005}
{Taylor}, M.~B. 2005, in Astronomical Society of the Pacific Conference Series,
  Vol. 347, Astronomical Data Analysis Software and Systems XIV, ed.
  P.~{Shopbell}, M.~{Britton}, \& R.~{Ebert}, 29

\bibitem[{{Torres} {et~al.}(2006){Torres}, {Quast}, {da Silva}, {de La Reza},
  {Melo}, \& {Sterzik}}]{Torres+06}
{Torres}, C.~A.~O., {Quast}, G.~R., {da Silva}, L., {et~al.} 2006, \aap, 460,
  695

\bibitem[{{Torres} {et~al.}(2008){Torres}, {Quast}, {Melo}, \&
  {Sterzik}}]{Torres2008}
{Torres}, C.~A.~O., {Quast}, G.~R., {Melo}, C.~H.~F., \& {Sterzik}, M.~F. 2008,
  {Young Nearby Loose Associations}, ed. B.~{Reipurth}, Vol.~5, 757

\bibitem[{{Ujjwal} {et~al.}(2020){Ujjwal}, {Kartha}, {Mathew}, {Manoj}, \&
  {Narang}}]{Ujjwal+20}
{Ujjwal}, K., {Kartha}, S.~S., {Mathew}, B., {Manoj}, P., \& {Narang}, M. 2020,
  arXiv e-prints, arXiv:2002.04801

\bibitem[{{White} {et~al.}(2007){White}, {Gabor}, \& {Hillenbrand}}]{White2007}
{White}, R.~J., {Gabor}, J.~M., \& {Hillenbrand}, L.~A. 2007, \aj, 133, 2524

\bibitem[{{Zuckerman} {et~al.}(2001){Zuckerman}, {Song}, {Bessell}, \&
  {Webb}}]{Zuckerman+01}
{Zuckerman}, B., {Song}, I., {Bessell}, M.~S., \& {Webb}, R.~A. 2001, \apjl,
  562, L87

\end{thebibliography}

\begin{appendix}

\section{Cross-match with \textit{Gaia} DR2}  
\label{App:match-Gaia}

In Section~\ref{sec:data}, we cross-matched our sample of candidate members of $\beta$~Pic with the \textit{Gaia} DR2 catalogue to obtain the proper motions and parallaxes. There are six sources which are not in the \textit{Gaia} DR2 catalogue and eight which only have the two-parameter solution in \textit{Gaia}. In this Appendix, we discuss the reasons for which these sources where not in the DR2 catalogue and the perspectives for eDR3, expected for the end of 2020.

There are two sources, \object{2MASS J05120636-2949540} and \object{2MASS J04210718-6306022}, with magnitudes $G>21$~mag which fail the first condition to have a five-parameter solution in \textit{Gaia}. The other six have an \texttt{astrometric\_sigma5d\_max} too large and fail the third condition (equation~11 from \citealt{Lindegren+2018}). The \texttt{astrometric\_sigma5d\_max} is a parameter used to detect cases where one or several parameters from the five-parameter solution are poorly determined. These stars are very nearby and have high proper motions which could hinder the proper cross-match of the observed transits. In addition, at least two are spectroscopic binaries (\object{2MASS J20100002-2801410} and \object{2MASS J21374019+0137137}), a fact that could difficult the derivation of a proper AGIS solution.

There are three sources, \object{2MASS J00160844-0043021}, \object{2MASS J03582255-4116060}, and \object{2MASS J23433470-3646021} with a magnitude $J\gtrsim15.8$~mag, which are fainter than the \textit{Gaia} detection limit. It is expected that they will not appear in any of the future \textit{Gaia} releases. We checked\footnote{\url{https://gaia.esac.esa.int/gost/}} that the other three sources, \object{2MASS J01112542+1526214}, \object{2MASS J03323578+2843554}, and \object{2MASS J05241914-1601153}, have a \texttt{visibility\_period\_used} $<6$ and for that reason were rejected from the five-parameter solution. In addition, they are known to be close spectroscopic binaries with separations of 0.2--0.6\arcsec which can induce to an erroneous solution and are not included in \textit{Gaia} DR2 \citep{Lindegren+2018}.

\section{Kinematically discarded sources}  
\label{App:discarded}

\begin{figure}
    \includegraphics[width = \columnwidth]{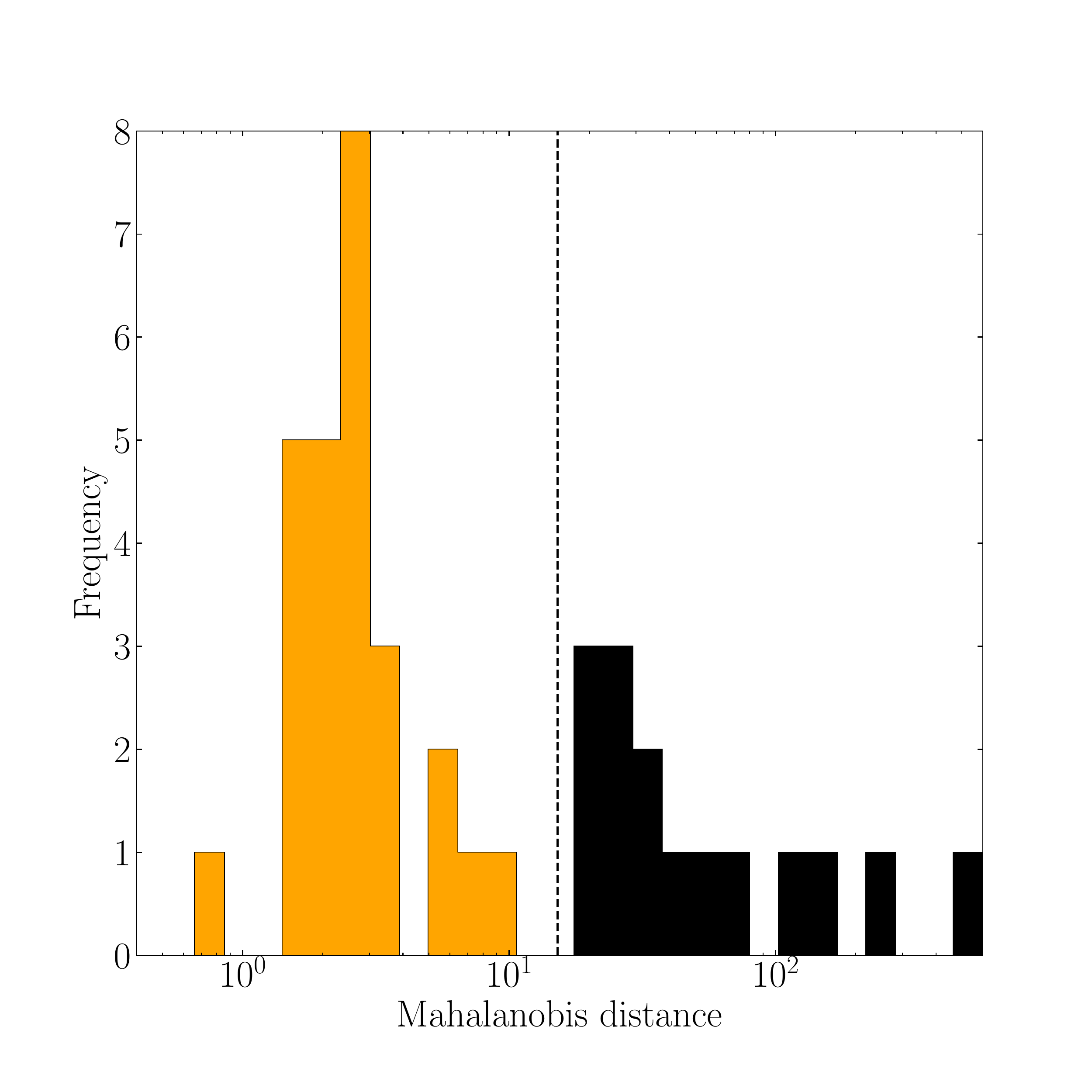}
    \caption{Histogram of the Mahalanobis distance to the centre of the velocity distribution ($\dot{\xi}^\prime$, $\dot{\eta}^\prime$, $\dot{\zeta}^\prime$) of the 42 single sources of our sample. The vertical dashed line indicates the percentile $p_{65}$ used to select the kinematic members (see Sect.~\ref{subsec:vel_dist}).
    }
    \label{fig:mahalanobis_dist_kin_selection}
\end{figure}{}

Here, we discuss possible reasons for which the 3D velocity of the 15 kinematic outliers reported in Sect~\ref{subsec:vel_dist} was found inconsistent with the rest of members of $\beta$~Pic. We also review the two suspected spectroscopic binaries found in this work (see Sect.~\ref{section2.3.2}) and the outlier in 3D positions (which was rejected because of the different orbital motion with respect to the $\beta$~Pic members, see Sect.~\ref{subsec:vel_dist}). 

\subsection*{\object{2MASS J01365516-0647379}}
This source was first classified as a candidate member of $\beta$~Pic by \citet{Malo2013} with a low membership probability of 27.4\%, taking into account the position, proper motion, magnitude, and colour. Later, \citet{Malo2014} revised the membership of this source and found a probability of 99.9\% including the radial velocity of \citet{Shkolnik2012}.
Our radial velocity estimate ($13.02\pm0.18$~km~s$^{-1}$) is consistent with the value of \citet{Shkolnik2012} ($12.2\pm0.4$~km~s$^{-1}$). Recently, \citet{Crundall+19} also classified this source as a field contaminant based on \textit{Gaia} DR2 astrometry and the radial velocity of \citet{Shkolnik2012}.

\subsection*{\object{2MASS J01373545-0645375}}
This source was proposed as a candidate of $\beta$~Pic by \citet{Gagne2018a}. However, it had been previously classified as a member of the Hercules Lyra association by \citet{Lopez-Santiago+06} and \citet{Gagne2018a} could not confirm its membership because they did not consider the Hercules Lyra association in their analysis. Our radial velocity estimate ($12.01\pm0.12$~km~s$^{-1}$) is similar to a recent value from the literature ($11.658\pm0.006$~km~s$^{-1}$, \citealt{Soubiran+18}).

\subsection*{\object{2MASS J02232663+2244069}}
Our radial velocity measurement ($12.60\pm0.15$~km~s$^{-1}$) is consistent with the one from the \textit{Gaia} DR2 catalogue ($12.1\pm0.6$~km~s$^{-1}$). This source was listed as a high probability (99\%) member of $\beta$~Pic by \citet{Malo2013} based on a radial velocity and a proper motion which differ by 2~km~s$^{-1}$ and 6 mas~yr$^{-1}$, respectively, from \textit{Gaia} DR2. The different data could explain why this source was discarded by our kinematic selection and the membership of this source has been revised with our data.

\subsection*{\object{2MASS J03573393+2445106}}
We have three spectra for this source with radial velocity measures of $13.46\pm0.18$~km~s$^{-1}$ (2018-08-12), $13.44\pm0.18$~km~s$^{-1}$ (2018-08-14), and $15.30\pm0.14$~km~s$^{-1}$ (2019-11-30). This source is rotationally variable (0.86~days, \citealt{Hartman+2011}) which could explain the variations in the radial velocity that we measure. This source is a candidate of spectroscopic binary which requires more follow-up observations to confirm it. We also note that \citet{Crundall+19} classified it as a field contaminant.

\subsection*{\object{2MASS J05004928+1527006}}
This source was classified as a member by \citet{Schlieder2012} based on a predicted radial velocity of $13.70\pm2.03$~km~s$^{-1}$. We measure a radial velocity of $18.4\pm0.3$~km~s$^{-1}$, similar to what has been reported in the literature \citep{White2007}, and significantly different to the predicted value used in the previous membership analysis. Additionally, this source has been classified as a member of the Taurus-Auriga complex \citep{Kraus+17}, and therefore is a likely contaminant in $\beta$~Pic.

\subsection*{\object{2MASS J08475676-7854532}}
This source was classified as a candidate of $\beta$~Pic based on a predicted radial velocity of $13.4\pm1.5$~km~s$^{-1}$ \citep{Malo2013}. This value is significantly different from our measurement of $23.1\pm0.3$~km~s$^{-1}$ and with the literature ($23.4\pm0.3$~km~s$^{-1}$ from \citealt{Malo2014}). Using \textit{Gaia}, it has been proposed as a member of $\eta$~Chamaeleontis \citep{Cantat-Gaudin+18}.

\subsection*{\object{2MASS J11493184-7851011}}
This source was classified as a $\beta$~Pic candidate based on a predicted radial velocity of $10.8\pm1.6$~km~s$^{-1}$ and a predicted distance of 68~pc ($\varpi=14.7$~mas) by \citet{Malo2014}. Our two radial velocity measurements differ by about 1.3~km~s$^{-1}$ between them but have a mean value of $14.5\pm0.8$~km~s$^{-1}$, which is not compatible with the predicted radial velocity in that study. The \textit{Gaia} DR2 parallax of this source is $9.92\pm0.03$~mas, indicating this source is probably a contaminant. A recent study classified this source as a $\epsilon$~Chamaeleontis \citep{Schneider+19}.

\subsection*{\object{2MASS J13545390-7121476}}
This source was classified as a candidate member of $\beta$~Pic by \citet{Malo2014} based on proper motions values which differ of about 20~mas~yr$^{-1}$ from the values of \textit{Gaia} DR2. This source is probably a contaminant.

\subsection*{\object{2MASS J19312434-2134226}}
Our radial velocity measurement ($-36.6\pm1.8$~km~s$^{-1}$) is not consistent with the literature (e.g. \citealt{Shkolnik2012} measured a radial velocity of $-26.0\pm1.8$~km~s$^{-1}$ and \citealt{Malo2014} $-25.6\pm1.5$~km~s$^{-1}$) with a difference of about 10~km~s$^{-1}$. We checked the CCF and there are hints it might be a spectroscopic binary.
In addition, a recent study classified this as a member of the Argus association \citep{Janson+2017}.

\subsection*{\object{2MASS J21212873-6655063}}
This source was classified by \citet{Malo2014} as a high probability (99.9\%) member of $\beta$~Pic. However, their analysis was based on pre-\textit{Gaia} astrometry and the proper motions they used differ about 20~mas~yr$^{-1}$ from the one of \textit{Gaia} DR2, indicating the membership should be revised.

\subsection*{\object{2MASS J23314492-0244395}}
This source was classified as a $\beta$~Pic candidate member by \citet{Malo2013}. However, their analysis was based on pre-\textit{Gaia} astrometry and the proper motions they used differ about 10~mas~yr$^{-1}$ to the ones of \textit{Gaia} DR2.

\subsection*{\object{2MASS J23512227+2344207}}

Our radial velocity measurement ($-1.0\pm0.3$~km~s$^{-1}$) differs by about 1~km~s$^{-1}$ from the measurement of \citet{Shkolnik2012} ($-2.1\pm0.5$~km~s$^{-1}$). \citet{Binks+16} provided another radial velocity measure for this source ($38.6\pm1.6$~km~s$^{-1}$), with a discrepancy of several tens of km~s$^{-1}$. Based on their measurement, they rejected this source as a $\beta$~Pic member and also suggested the possibility of a binary system to explain the differences observed. \citet{Messina+16} classified this source as a single star based on a study of photometric variability. Further work is required to confirm the binarity of this source. Additionally, other authors have classified this source as member of other moving groups (e.g. \citealt{Shkolnik2012}, \citealt{Klutsch+2014}).

\subsection*{\object{2MASS J21183375+3014346}}
This source was classified as a candidate member of $\beta$~Pic by \citet{Schlieder2012} with a predicted radial velocity of $-15.1\pm0.9$~km~s$^{-1}$. This value is significantly different from our radial velocity measurement ($-22.0\pm0.3$~km~s$^{-1}$). Additionally, \citet{Shkolnik+17} recently measured a radial velocity similar to ours ($-22.5\pm0.8$~km~s$^{-1}$) and rejected the $\beta$~Pic membership of this source.

\subsection*{\object{2MASS J22571130+3639451}}
This source was classified as a candidate member of $\beta$~Pic by \citet{Schlieder2012} with a predicted radial velocity of $-10.0\pm0.9$~km~s$^{-1}$ although their measured radial velocity was $-20\pm1.2$~km~s$^{-1}$. We have analysed eight spectra of this source and obtained a variable radial velocity between $-10$~km~s$^{-1}$ and $-20$~km~s$^{-1}$, indicating this is probably an unresolved spectroscopic binary.

\subsection*{\object{2MASS J16120516-4556242}, \object{2MASS J17092947-5235197}, \object{2MASS J18430597-4058047}, and \object{2MASS J20105054-3844326}}
These sources were classified as new members of the $\beta$~Pic moving group by \citet{Gagne2018b} with no radial velocity measurements. The first estimation of their radial velocity provided for the first time in the present work, shows that their velocity is not compatible with the velocity distribution of $\beta$~Pic, suggesting they might be contaminants.

\section{Additional Tables and Figures}

\begin{figure}
    \includegraphics[width = \columnwidth]{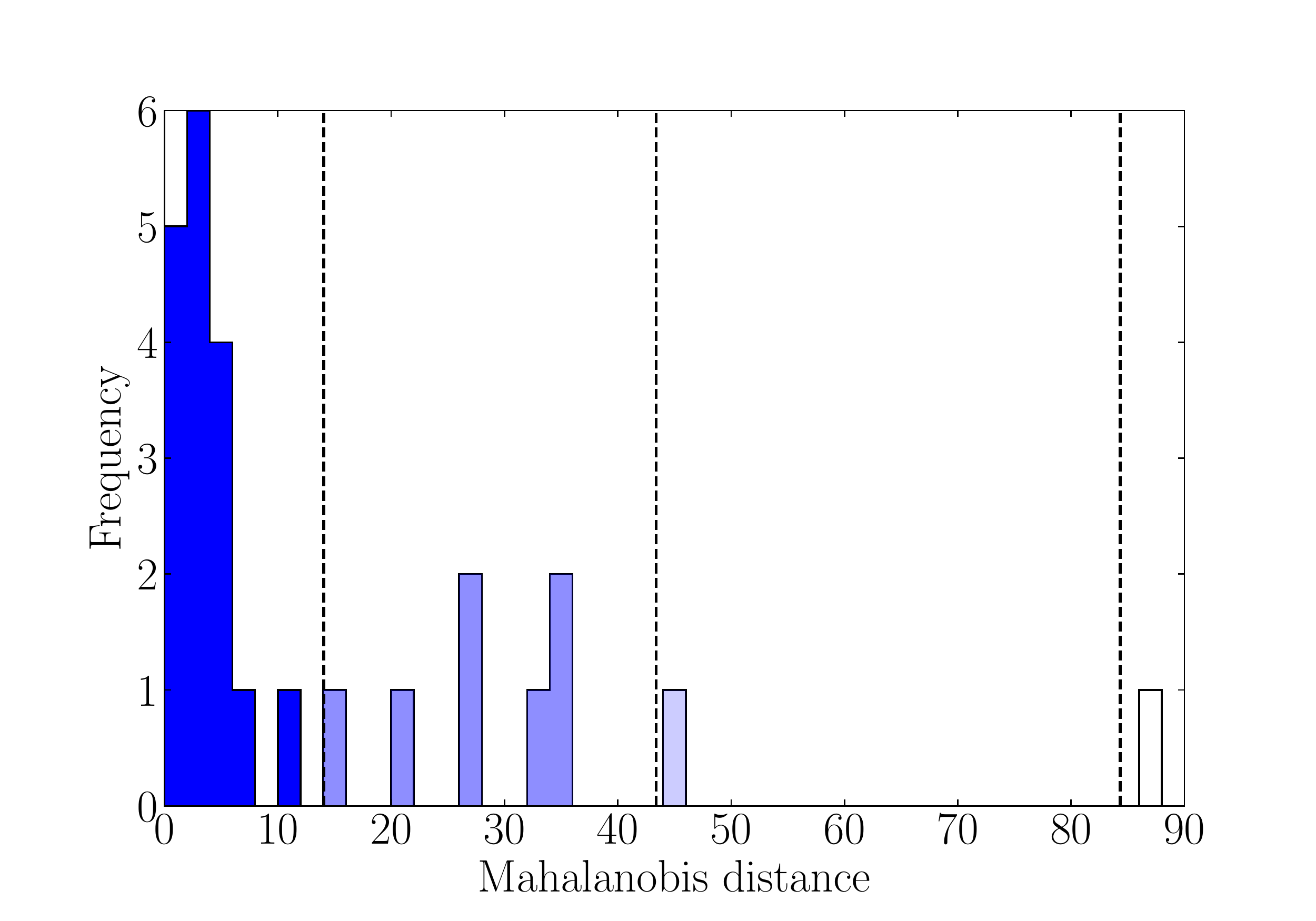}
    \caption{Histogram of the Mahalanobis distance to the centre of the 3D positions distribution ($\xi^\prime$, $\eta^\prime$, $\zeta^\prime$) of the 26 selected kinematic members of our sample, computed with the robust metric. The vertical dashed lines indicate the position of the percentiles $p_{68}$, $p_{95}$, and $p_{99.7}$.
    }
    \label{fig:mahalanobis_dist_core}
\end{figure}{}

\end{appendix}

\end{document}